\documentclass[a4paper,fleqn,usenatbib]{mnras}

\usepackage{newtxtext,newtxmath}
\usepackage[T1]{fontenc}
\usepackage{ae,aecompl}

\usepackage{natbib}

\usepackage{graphicx}
\usepackage{amsmath}	
\usepackage{amssymb}	
\usepackage{mathrsfs}
\usepackage{url}
\usepackage{bm}
\usepackage{booktabs}
\usepackage{multicol}
\usepackage{comment}

\def\gtsima{$\; \buildrel > \over \sim \;$}
\def\ltsima{$\; \buildrel < \over \sim \;$}
\def\gtrsim{\lower.5ex\hbox{\gtsima}}
\def\lesssim{\lower.5ex\hbox{\ltsima}}

\newcommand{\diff}[2]{\frac{\mathrm{d} #1}{\mathrm{d #2}}}

\newcommand{\msun}{\,\mathrm{M}_{\sun{}}}
\newcommand{\zsun}{\,Z_{\sun{}}}
\newcommand{\rsun}{\,\mathrm{R}_{\sun{}}}

\newcommand{\sevn}{\textsc{SEVN}}
\newcommand{\mobse}{\textsc{MOBSE}}
\newcommand{\bse}{\textsc{BSE}}
\newcommand{\parsec}{\textsc{PARSEC}}
\newcommand{\combine}{\textsc{ComBinE}}

\begin{document}
\title[Merging black hole binaries with the SEVN code]{Merging black hole binaries with the SEVN code}
\author[Spera et al.]{
Mario Spera$^{1,2,3,4,5,6}$\thanks{E-mail: mario.spera@live.it}, 
Michela Mapelli$^{1,2,3,4}$\thanks{E-mail: michela.mapelli@oapd.inaf.it}, 
Nicola Giacobbo$^{1,2,3}$, 
\newauthor
Alessandro A. Trani$^{3,7,8}$, 
Alessandro Bressan$^{3,8}$, 
and Guglielmo Costa$^{8}$\\
$^1$Dipartimento di Fisica e Astronomia `G. Galilei', University of Padova, Vicolo dell'Osservatorio 3, I--35122, Padova, Italy\\
$^2$INFN, Sezione di Padova, Via Marzolo 8, I--35131, Padova, Italy\\
$^3$INAF, Osservatorio Astronomico di Padova, Vicolo dell'Osservatorio 5, I--35122, Padova, Italy\\
$^4$Institut f\"ur Astro- und Teilchenphysik, Universit\"at Innsbruck, Technikerstrasse 25/8, A-6020, Innsbruck, Austria\\
$^5$ Department of Physics and Astronomy, Northwestern University, Evanston, IL 60208, USA\\
$^6$Center for Interdisciplinary Exploration and Research in Astrophysics (CIERA), Evanston, IL 60208, USA\\
$^7$Department of Astronomy, Graduate School of Science, The University of Tokyo, 7-3-1 Hongo, Bunkyo-ku, Tokyo, 113-0033, Japan\\
$^8$SISSA, via Bonomea 265, I-34136 Trieste, Italy
}

\maketitle \vspace {7cm }
\bibliographystyle{mnras}
 
\begin{abstract}
	Studying the formation and evolution of black hole binaries (BHBs) is essential for the interpretation of current and forthcoming gravitational wave (GW) detections. We investigate the statistics of BHBs that form from isolated binaries, by means of a new version of the \sevn{} population-synthesis code. \sevn{} integrates stellar evolution by interpolation over a grid of stellar evolution tracks. We upgraded  \sevn{}  to include binary stellar evolution processes and we used it to evolve a sample of $1.5\times{}10^8$ binary systems, with metallicity in the range $\left[10^{-4};4\times 10^{-2}\right]$. From our simulations, we find that the mass distribution of black holes (BHs) in double compact-object binaries is remarkably similar to the one obtained considering only single stellar evolution. The maximum BH mass we obtain is $\sim 30$, $45$ and $55\msun{}$ at metallicity $Z=2\times 10^{-2}$, $6\times 10^{-3}$, and $10^{-4}$, respectively. A few massive single BHs may also form ($\lesssim 0.1\%$ of the total number of BHs), with mass up to $\sim 65$, $90$ and $145\msun{}$ at $Z=2\times 10^{-2}$, $6\times 10^{-3}$, and $10^{-4}$, respectively. These BHs fall in the mass gap predicted from pair-instability supernovae. We also show that the most massive BHBs are unlikely to merge within a Hubble time.  In our simulations, merging BHs like {   GW151226 and GW170608, form at all metallicities, the high-mass systems (like GW150914, GW170814 and GW170104) originate from metal poor ($Z\lesssim{}6\times 10^{-3}$) progenitors, whereas GW170729-like systems are hard to form, even at $Z = 10^{-4}$}. The BHB merger rate in the local Universe obtained from our simulations is  $\sim 90 \mathrm{Gpc}^{-3}\mathrm{yr}^{-1}$,  consistent with the rate inferred from LIGO-Virgo data.
\end{abstract}

%

\section{Introduction}
The existence of double black hole binaries (BHBs) has been hypothesized  for several decades \citep{tutukov1973,thorne1987,schutz1989,kulkarni1993,sigurdsson1993,portegieszwart2000,colpi2003,belczynski2004}, but their first observational confirmation is the detection of GW150914 in September 2015 \citep{abbott2016a}. Since then, {   nine} additional BHB mergers have been reported by the LIGO-Virgo collaboration \citep{LIGOdetector,Virgodetector}: {   GW151012 \citep{catalog2018}, GW151226 \citep{abbott2016b}, GW170104 \citep{abbott2017a}, GW170608 \citep{abbott2017b}, GW170729, GW170809 \citep{catalog2018}, GW170814 \citep{abbott2017c}, GW170818, and GW170823 \citep{catalog2018}.}

{   Seven} of the observed merging systems host black holes (BHs) with mass larger than $\sim{}30$ M$_\odot$. These massive BHs were a surprise for the astrophysics community, because there is no conclusive evidence for BHs with mass $>20$ M$_\odot$ from X-ray binaries\footnote{The compact object in the X-ray binary IC10 X-1 was estimated to have a mass of $\sim{}28-34$ M$_\odot$ \citep{prestwich2007,silverman2008}, but this result is still debated \citep{laycock2015}.} \citep{ozel2010,farr2011}.

If these BHs formed from the collapse of massive stars, such large masses require the progenitors to be massive metal-poor stars \citep{mapelli2009,mapelli2010,belczynski2010,mapelli2013,mapelli2014,spera2015}. Massive metal-poor stars are thought to lose less mass by stellar winds than their metal-rich analogues \citep{vink2001,graefener2008,vink2011}. Thus, a metal-poor star ends its life with a larger mass than a metal-rich star with the same zero-age main sequence (ZAMS) mass. Although our knowledge of the hydrodynamics of core-collapse supernovae (SNe) is far from optimal  (see \citealt{foglizzo2015} for a recent review), several studies \citep{fryer1999,fryer2001,heger2003,oconnor2011,fryer2012,ugliano2012,ertl2016} suggest that if the mass and/or the compactness of the star at the onset of collapse are sufficiently large, then the star can avoid a SN explosion and collapse to a BH promptly, leading to the formation of a relatively massive BH. Since metal-poor stars lose less mass by stellar winds, they are also more likely to form massive BHs via direct collapse than metal-rich stars  \citep{spera2015,belczynski2016}. Stellar rotation (e.g. \citealt{limongi2017,limongi2018}), magnetic fields \citep{petit2017}, pair-instability SNe (PISNe) and pulsational pair-instability SNe (PPISNe) \citep{belczynski2016pair,spera2017,woosley2017} also affect this picture.

Other possible scenarios for the formation of $\sim{}30-40$ M$_\odot$ BHs include primordial BHs (i.e. BHs formed by gravitational instabilities in the very early Universe, e.g. \citealt{carr2016, sasaki2018}) and second-generation BHs (i.e. BHs formed from the mergers of smaller stellar BHs, \citealt{gerosa2017}). Stellar dynamics in dense star clusters can also affect the final mass of merging BHs (e.g. \citealt{portegieszwart2004,giersz2015,mapelli2016}). 

Overall, the formation of massive stellar BHs ($30-40$ M$_\odot$) is still an open question, several aspects of massive star evolution and core-collapse SN explosions being poorly understood. 

The formation channels of BHBs are even more debated. A BHB can form from the evolution of massive close stellar binaries (e.g. \citealt{tutukov1973,bethe1998,belczynski2016,demink2016,mandel2016,marchant2016,mapelli2018,giacobbo2018}) or from dynamical processes involving BHs in dense star clusters (e.g. \citealt{portegieszwart2000,colpi2003,ziosi2014,giersz2015,kimpson2016,mapelli2016,askar2016,rodriguez2016,banerjee2017}). In this manuscript, we will focus on the evolution of a massive close stellar binary in ``isolation'', that is without considering dynamical processes in star clusters. 

A large fraction of massive stars ($\sim{}50-70$ \%, \citealt{sana2012}) are members of binary systems since their birth. The evolution of  a close stellar binary is affected by a number of physical processes, such as mass transfer (via stellar winds or Roche lobe overflow), common envelope (CE) and tides (e.g. \citealt{portegieszwart1996,bethe1998,hurley2002}). Thus, the final fate of a binary member can be completely different from that of a single star with the same ZAMS mass and metallicity. This affects the statistics of merging BHBs, because it changes the number of BHBs and their properties (masses, eccentricities, semi-major axes and spins).

Binary population-synthesis codes have been used to study the evolution of massive binaries and their impact for the demography of BHBs. Since the pioneering work by \cite{whyte1985}, several population-synthesis codes have been developed. The `binary-star evolution' ({\sc BSE}) code \citep{hurley2000,hurley2002} is surely one of the most used population-synthesis codes. Stellar evolution is implemented in {\sc BSE} through polynomial fitting formulas, making this code amazingly fast. 
The fitting formulas adopted in {\sc BSE} are based on quite outdated stellar evolution models.  For this reason, \cite{giacobbo2018} and \cite{giacobbo2018b} have updated the recipes for stellar winds and SN explosions in {\sc BSE}, producing a new  version of {\sc BSE} called `Massive Objects in Binary Stellar Evolution' ({\sc MOBSE}). 

Many other population-synthesis codes are based on updated versions of \cite{hurley2000} fitting formulas, including {\sc SeBa} \citep{portegieszwart1996,toonen2012,mapelli2013,schneider2017}, {\sc binary\_c} \citep{izzard2004,izzard2006,izzard2009}, {\sc StarTrack} \citep{belczynski2008,belczynski2010,belczynski2016} and {\sc COMPAS} \citep{stevenson2017,barrett2017}.

Alternative approaches to fitting formulas consist in integrating stellar evolution on the fly (e.g. {\sc BPASS}, \citealt{eldridge2016,eldridge2017}; {\sc MESA}, \citealt{paxton2011,paxton2013,paxton2015}) or in  reading stellar evolution from look-up tables (e.g. {\sc SEVN} \citealt{spera2015,spera2017}; {\sc ComBinE}, \citealt{kruckow2018}). The interpolation of stellar evolution from look-up tables, containing a grid of stellar evolution models, is both convenient in terms of computing time and versatile, because the stellar evolution model can be updated by simply changing tables.

In this manuscript, we discuss the statistics of BHBs we obtained with the {\sc SEVN} code \citep{spera2015,spera2017}. {\sc SEVN} interpolates stellar evolution from look-up tables (the default tables being derived from {\sc PARSEC}, \citealt{bressan2012,chen2015}), includes five different models for core-collapse SNe, contains prescriptions for PPISNe and PISNe and has been updated to implement also binary evolution processes (wind mass transfer, Roche lobe overflow, CE, stellar mergers, tidal evolution, gravitational wave decay and magnetic braking).


\section{The SEVN code}\label{sec:sevn}
\subsection{Single star evolution} 
\subsubsection{Interpolation method}\label{sec:interp}
\sevn{} evolves the physical parameters of stars by reading a set of tabulated stellar evolutionary tracks that are interpolated on-the-fly. As default, \sevn{} includes a new set of look-up tables generated using the {\sc PARSEC} code \citep{bressan2012,chen2014,tang2014,chen2015}. This set of tables ranges from metallicity $Z=10^{-4}$ to $Z=6\times 10^{-2}$ with stars in the mass range $2\leq M_{\mathrm{ZAMS}}/\msun{}\leq 150$. Furthermore, we have used the \parsec{} code to generate a new set of tracks for bare Helium cores to follow the evolution of the stars that lose the whole Hydrogen envelope after a mass-transfer phase. The look-up tables of Helium stars range from metallicity $Z=10^{-4}$ to $Z=5\times 10^{-2}$ with stars in the mass range $0.4\leq M_{\mathrm{He-ZAMS}}/\msun{}\leq 150$ (see Sec. \ref{sec:purehe} for details).


To perform the interpolation, in \sevn{} we distinguish the stars that are on the main sequence (H phase) from those that have already formed a He core but not yet a Carbon-Oxygen (CO) core (He phase) and those that have already formed a CO core (CO phase). The division into three macro-phases is convenient in terms of computing time and it also ensures that the stars used for the interpolation have the same internal structure. Furthermore, we impose that the interpolating stars have the same percentage of life ($\Theta_{\rm p}$) of the interpolated star on its macro-phase.  
For every time $t$, the percentage of life of a star is
\begin{equation}
\Theta_{\rm p} = \frac{t - t_{0,\mathrm{p}}}{t_{\mathrm{f,p}} - t_{0,\mathrm{p}}}
\label{eq:percentlife}
\end{equation}
where $t_{0,\mathrm{p}}$ is the starting time of the star's evolutionary macro-phase $p$ (where $p={\rm H}$ phase, He phase and CO phase) and $t_{\mathrm{f,p}}$ is its final time. 
By using $\Theta_{\rm p}$, we ensure that the stars used for the interpolation are at the same stellar evolutionary phase within the same macro-phase.    

In addition to these three macro-phases, we have defined several stellar-evolution phases. As in \citet{hurley2002}, in \sevn{} we use integer values to distinguish between different stellar-evolution phases. Table \ref{tab:table1} shows the list of the stellar evolutionary phases and their corresponding macro-phases used in the \sevn{} code. We adopt the same indexes used by \citet{hurley2002} except for the massless remnants for which we use the index $-1$ instead of $15$. The stellar evolution phase of a star is evaluated using the values and the rate of change of the interpolated physical stellar parameters. It is worth noting that in \sevn{} we mark a star as Wolf-Rayet (WR, $k=7,8,9$) if 

\begin{equation}
\frac{\left|M-M_{\mathrm{He}}\right|}{M}<2\times 10^{-2},
\label{eq:WRcond}
\end{equation}

where $M$ is the total mass of the star and $M_{\mathrm{He}}$ is its He-core mass.
The details of the interpolation method for isolated stars are discussed in the supplementary material, Appendix A1.


While for isolated stars the interpolation tracks are fixed, for binary stars we allow jumps on different tracks. Every time a star has accreted (donated) a significant amount of mass $\Delta m$ from (to) its companion, the \sevn{} code moves onto another evolutionary track in the look-up tables. The value of $\Delta m$ depends on the binary evolution processes (see Sec. \ref{sec:bse}) but we allow jumps to new tracks only if 

\begin{equation}
  \Delta m>\gamma_{m}M,
\label{eq:chtracktext}
\end{equation}

where $M$ is the total mass of the star and $\gamma_{m}$ is a parameter with typical value of $\sim 0.01$. 

The jumps onto new tracks depend primarily on the star's macro-phase. For a star in the H phase, we search for new interpolating stars with (i) $t<t_{\mathrm{f,H\,{}phase}}$, (ii) the same percentage of life of the star, and (iii) the same total mass.

For a star in the He phase, the interpolating stars must have $t>t_{0,\mathrm{He\,{}phase}}$ and the same He core mass. If the interpolated star is not a WR star, we also impose that the new track has the same mass of the H envelope.

For stars in the CO phase, we use the same strategy adopted for stars in the He phase but we require that $t>t_{0,\mathrm{CO\,{}phase}}$. In all cases, if the requirements are not matched, we use the best interpolating stars the algorithm was able to find. The details of the track-finding method are discussed in the supplementary material, Appendix A2.

\begin{table}
\begin{center}
 \leavevmode
\begin{tabular}[!h]{ccc}
\hline
$k$ &  Phase   & Macro-phase \\
\hline
0 & Low-mass main sequence (MS, $M<0.7\msun{}$)   & H phase\\
1 & MS ($M>0.7\msun{}$) & H phase\\
2 & Hertzsprung gap (HG) & He phase\\
3 & First giant branch & He phase\\
4 & Core He burning & He phase\\
5 & Early asymptotic giant branch (AGB) & CO phase\\
6 & Thermally pulsing AGB & CO phase\\
7 & Naked Helium MS & He phase\\
8 & Naked Helium HG & CO phase \\
9 & Naked Helium giant branch & CO phase\\
10 & He white dwarf (WD) & none\\
11 & Carbon-Oxygen WD & none\\
12 & Oxygen-Neon WD & none\\
13 & Neutron star (NS) & none\\
14 & Black hole (BH) & none\\
-1 & Massless remnant & none\\         
\noalign{\vspace{0.1cm}}
\hline
\end{tabular}
\caption{\label{tab:table1} List of the integer values $k$ used for stellar evolutionary phases and their corresponding macro-phases. A naked Helium MS is a naked Helium star burning Helium in the core. A naked Helium HG is a naked Helium star burning Helium in shells. A naked Helium giant branch is a naked Helium star burning Carbon (or a heavier element) in the core.}
\end{center}
\end{table}


\subsubsection{Helium stars}
\label{sec:purehe}
The evolution  of the He stars is computed starting from a Helium ZAMS (He-ZAMS) obtained by removing the H-rich envelope of a normal star at the beginning of the central
He-burning phase and, thereafter, varying its total mass keeping the chemistry fixed.

The initial mass on the He-ZAMS varies from $0.36\msun{}$ to $150\msun{}$ with increasing mass steps of 0.02, 0.05, 0.10, 0.20, 0.50,  1.0,  2.0, 5.0, 20.0$\msun{}$ respectively above 0.36,  0.5, 0.8,  2,  9, 12, 20, 40, 100$\msun{}$.

The basic input physics is the same as that described in  ~\cite{bressan2012} and \cite{chen2015},
a part from the following small changes.
The nuclear reaction rates from the JINA REACLIB database \citep{cyburt2010} have been updated to their recommended values of April 6, 2015 \citep{Fu2018}.
The equation of state for He and and CO rich mixtures has been extended to slightly lower temperatures, as well as the corresponding radiative opacities.
We account for mass loss adopting the same mass-loss rates used for the PARSEC evolutionary tracks of massive stars in the WR phases \citep{chen2015}.

\begin{figure}
	\includegraphics[width=\columnwidth]{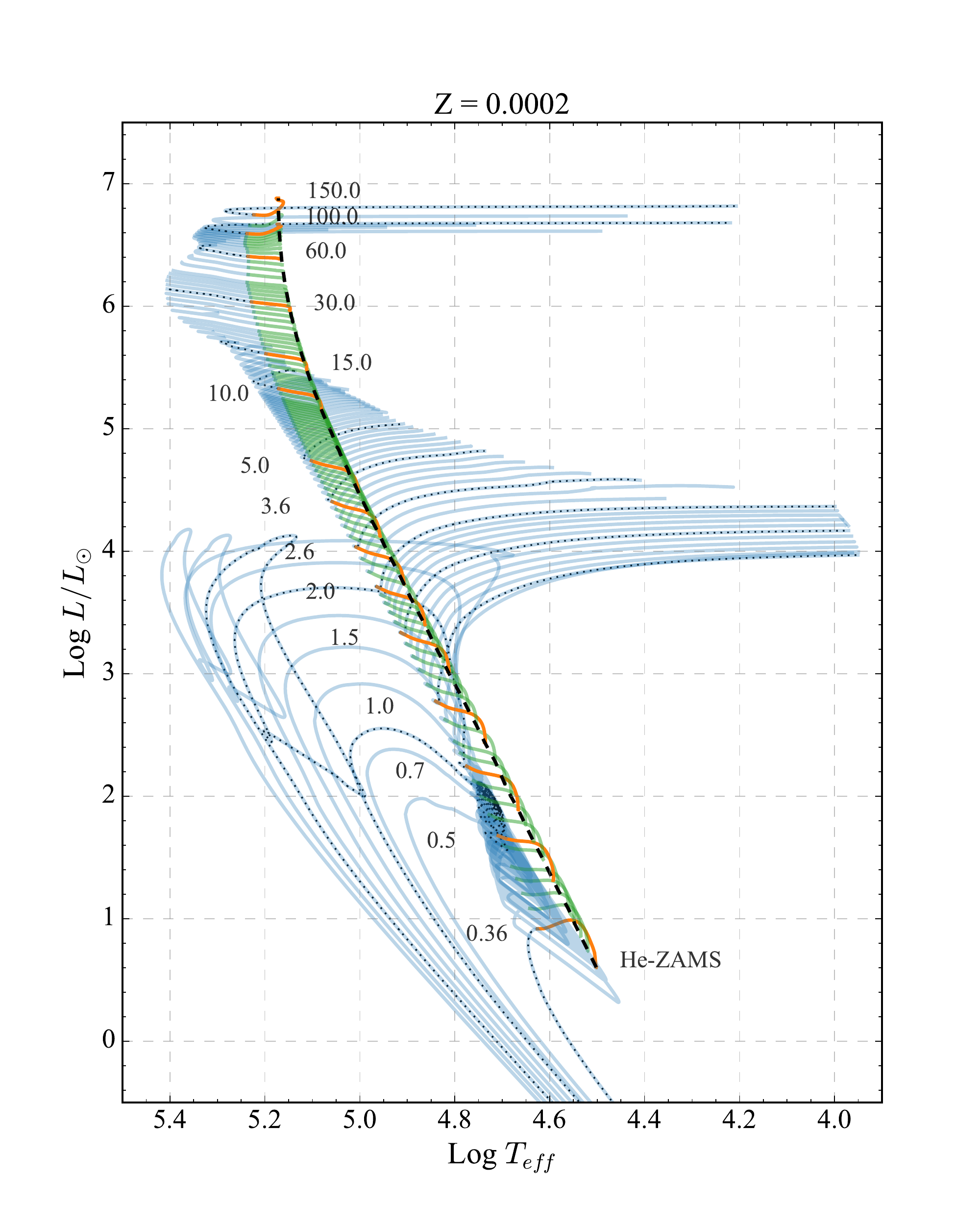}
	\caption{Hertzsprung-Russell (HR) diagram of the pure He-star tracks, at $Z = 2\times 10^{-4}$.
		The He-ZAMS is indicated by the black dashed line. The the central He-Burning phase is plotted in green (in orange for the labelled masses) to better show the width of the most populated area in the HR diagram. The remaining evolution (post He-MS) is coloured in blue.
		\label{fig:HR_He_lowz}}
\end{figure}

\begin{figure}
	\includegraphics[width=\columnwidth]{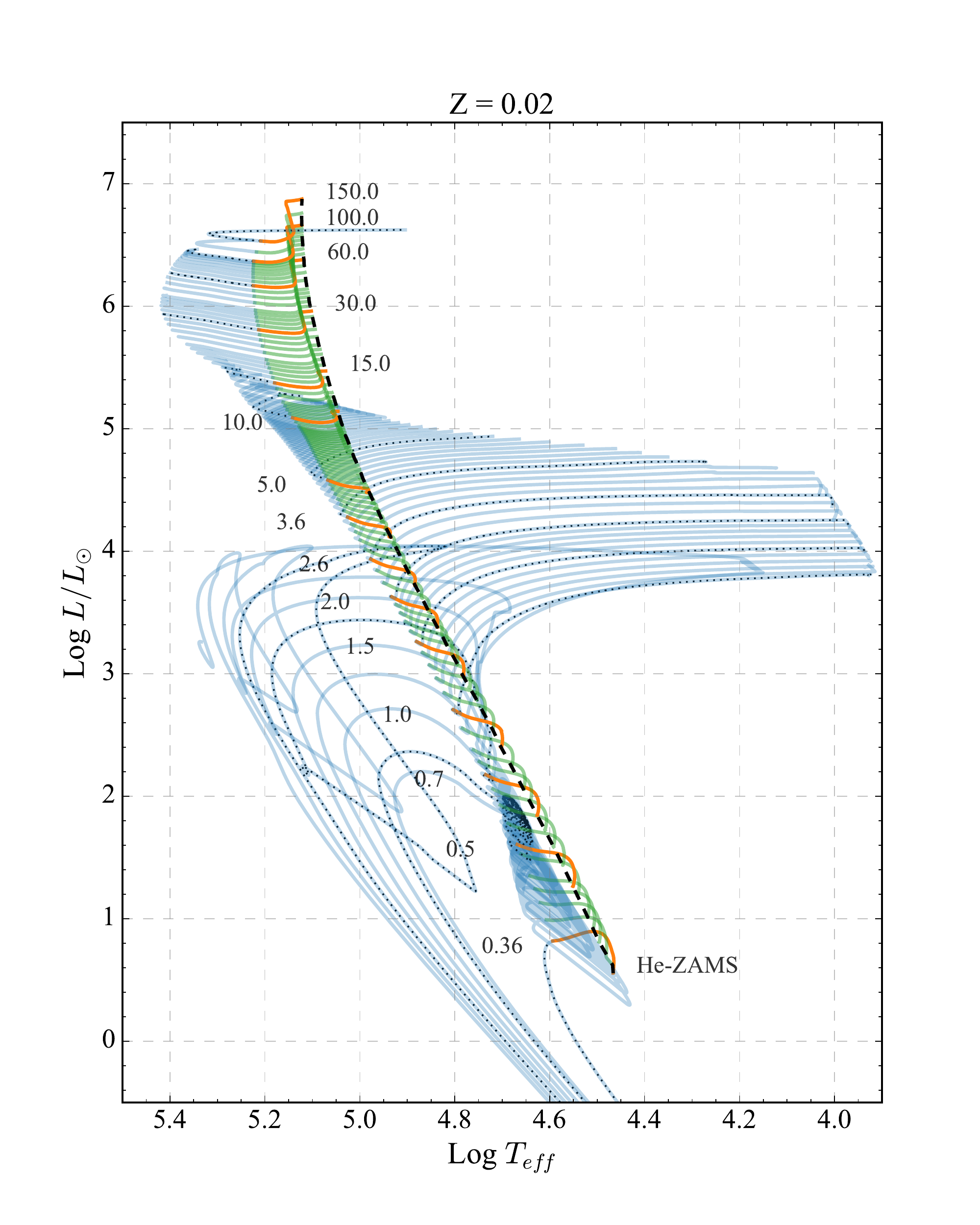}
	\caption{Same as Fig. \ref{fig:HR_He_highz} but for $Z = 2\times 10^{-2}$.}
	\label{fig:HR_He_highz}
\end{figure}

The evolution of selected sets of  naked He-star models is shown in
Figure~\ref{fig:HR_He_lowz} and Figure~\ref{fig:HR_He_highz}  for $Z=2\times 10^{-4}$ and $Z=2\times 10^{-2}$, respectively.
Here we briefly describe the evolution of the He stars with solar metallicity leaving a more thorough discussion to a companion paper.

The evolution on the Helium main sequence (He-MS) is very similar for all masses and characterized by a growing temperature  as the central He is burned.
At central He exhaustion the evolution is reversed and the stars move toward the asymptotic giant branch (AGB) or red super-giant  branch (RSGB), at least for initial masses below about 15~M$_\odot$.
For the stars with the lower masses ($0.36\msun{}$ to $0.9\msun{}$) the mass-loss is high enough to remove the surrounding  He-rich envelope before they reach the AGB and they evolve along the so called {\sl AGB-manqu\'e} phase and cool down along the CO-rich white dwarf (WD) sequence (see the tracks of the models with  $M_{\rm He-ZAMS}=0.5$, 0.6 and 0.8~M$_\odot$).

As in the case of the low-mass H-rich stars, the post-AGB phase is faster at increasing mass.

He stars with mass between 1.0~M$_\odot$ to 2.4~M$_\odot$ evolve toward the AGB branch
and the equation of state in their central regions begins to be  dominated  by degenerate electrons.
Stars with initial mass below 1.4~M$_\odot$  end their lives as CO WDs
because mass loss is able to decrease their current mass
below the threshold for Carbon ignition.
He stars with initial mass between 1.4~M$_\odot$
and to 2.4~M$_\odot$ could still ignite Carbon while not having a strongly degenerate electron core.
To better understand the evolution of these stars, we have followed in more
detail the evolution of He stars with initial mass between
1.5~M$_\odot$ to 2.4~M$_\odot$.
For $Z=0.0002$ we find that stars with $M_{\rm He-ZAMS}$ between 1.8~M$_\odot$ to 2.2~M$_\odot$
are able to ignite Carbon and, through  a series of off center Carbon burning episodes,
they build up a degenerate Oxygen-Neon-Magnesium core. These stars become  super AGB stars and their following fate is then dictated by the competition between the core growth by the Helium/Carbon burning shells and the envelope consumption by mass loss.

If the mass-loss process is high enough  to prevent the core mass to reach
the threshold density for the onset of electron-capture processes
on $^{24}$Mg and $^{20}$Ne nuclei, then
the star will become an Oxygen-Neon-Magnesium WD. Alternatively the star
will end its life as an electron-capture SN.
The threshold core mass is confined between  $M_{\rm CO}\simeq{}1.38$ M$_\odot$ \citep{Miyaji1980} and $M_{\rm CO}\simeq{}1.37$ M$_\odot$ \citep{Nomoto1984,Takahashi2013}. 
The model with He-ZAMS mass $M_{\rm He-ZAMS}=1.8$ M$_\odot$ is evolved until its total mass is $M=1.3$ M$_\odot$
and the core mass is $M_{\rm CO}\sim{}1.095$~M$_\odot$. For the track with $M_{\rm He-ZAMS}=1.9$ M$_\odot$ the last computed model has a total mass of $M=1.2$ M$_\odot$ and a core mass of $M_{\rm CO}\sim{}1.179$~M$_\odot$, while for the  $M_{\rm He-ZAMS}=2.0$ M$_\odot$ track the last computed model has a total mass of $M=1.2$ M$_\odot$ and a core mass of $M_{\rm CO}\sim{}1.218$~M$_\odot$
These three models will become O-Ne-Mg WDs.

The model with $M_{\rm He-ZAMS}=2.2$~M$_\odot$ is followed until the current mass and the core mass
are $M=1.454$ M$_\odot$ and $M_{\rm CO}=1.301$ M$_\odot$, respectively.
The central density at this stage is $\rho_c=4.40\times{}10^8$~g~cm$^{-3}$, while the central and the off center temperatures are $T_{c}=2.39\times{}10^8$~K and $T_{\rm max}=6.34\times{}10^8$~K, respectively.
This star has almost reached the mass threshold for the ignition of Neon in a electron degenerate gas, but we cannot follow this phase because our network does not yet include electron-capture reactions. A simple extrapolation indicates that with the current mass-loss
and core-growth rates, a $\sim{}0.084$~M$_\odot$ envelope can be lost before the core reaches the critical mass for Neon ignition ($\sim{}0.0004$~Myr is the time required for the former against $\sim{}0.0179$~Myr for the latter). Thus this mass could be the separation mass between
O-Ne-Mg WDs and electron-capture SNe.

The model with $M_{\rm He-ZAMS}=2.4$~M$_\odot$ is followed until the central density reaches $\rho_c=2.41\times{}10^8$ g cm$^{-3}$ and the central temperature is T$_c=5.39\times{}10^8$~K. At this point the core mass is $M_{\rm CO}\sim{}1.39$~M$_\odot$. The star has a total mass  of $M=2.28$~M$_\odot$ and an off center maximum temperature of $T_{\rm max}=1.87~10^9$~K. 
In the off center region near the maximum temperature Neon has been almost completely burned
and, given the high central density and degeneracy, it is likely that the core will soon begin
the electron-capture collapse. 
Similar properties are found for models with $Z=0.02$. The track with mass $M_{\rm He-ZAMS}=2.4$~M$_\odot$ is followed
until the central density reaches $\rho_c=1.860\times{}10^8$ g cm$^{-3}$, with a central temperature of $T_c$~3.645~10$^8$~K and an off center maximum temperature of $T_{\rm max}$=9.420~10$^8$~K.
At this stage the core mass is $M_{\rm CO}\sim{}1.301$~M$_\odot$ and the total mass is
$M_{\rm cur}\sim$1.865~M$_\odot$. An extrapolation adopting the current mass-loss rate and He-core growth rate indicates that the model will reach the critical core mass for Neon ignition
about ten times faster than what required by mass loss to peal off the envelope to below the same limit. In contrast, the opposite occurs for the model of initial mass $M_{\rm He-ZAMS}=2.2$~M$_\odot$.

More massive stars are evolved until the beginning of Oxygen burning.

\subsubsection{Stellar spin}

We follow the evolution of stellar spin $\Omega_{\rm spin}$ by taking into account the change of moment of inertia, mass loss by stellar winds, magnetic braking and mass transfer. We compute the moment of inertia as in \cite{hurley2000}: 
\begin{equation}
I = 0.1\,M_{\rm c} \,R^2_{\rm c} + 0.21 \, (M-M_{\rm c})\,{} R^2
\end{equation}
where $M$ and $R$ are the stellar mass and radius, while $M_{\rm c}$ and $R_{\rm c}$ are the core mass and radius.

We assume that stellar winds carry away spin angular momentum uniformly from a thin shell at the stellar surface. We include spin down by magnetic braking for giant stars with convective envelopes (type $k=2$ to $6$, see eq. 111 of \citealt{hurley2000}).


In the present work, we neglect the effect of stellar spin on wind mass loss. The enhancement of stellar winds due to rotation will be investigated in a forthcoming work.

We evolve stellar spins even if the {\sc{PARSEC}} stellar tracks we use in this paper are calculated for non rotating stars. Although not fully consistent, this approach has been followed in the past by most population-synthesis codes, to enable the calculation of tidal forces and other spin-dependent binary evolution processes. In future works, we will include rotating stellar evolutionary tracks from Costa et al. (in preparation).

\subsection{Prescriptions for supernovae (SNe)} 
The prescriptions for SNe adopted in {\sc SEVN} were already described in \cite{spera2015} and in \cite{spera2017}. Herebelow we briefly summarize the most important features, while we refer the reader to the supplementary material, Appendix B,  for more details. 

{\sc SEVN} contains five different models for core-collapse SNe,  which can be activated with a different option in the parameter file. These are (i) the rapid core-collapse model \citep{fryer2012}, (ii) the delayed core-collapse model \citep{fryer2012}, (iii) the  prescription implemented in the {\sc{STARTRACK}} code \citep{belczynski2010}, (iv) a model based on the compactness parameter \citep{oconnor2011}, and the (v) two-parameter criterion by \cite{ertl2016}. In this paper, we adopt the rapid core-collapse SN model as the reference model.

PISNe and PPISNe are also included in {\sc{SEVN}} following the prescriptions discussed in \cite{spera2017}.

Finally, the SN kicks are implemented in {\sc{SEVN}} adopting the \cite{hobbs2005} kick distribution for both neutron stars (NSs) and BHs but
we scale the kick by the amount of fallback \citep{fryer2012}:
\begin{equation}
	V_{\rm kick} = (1 - f_{\rm fb})\,{}W_{\rm kick},
\end{equation}
where $f_{\rm fb}$ is the fallback factor (the explicit expression can be found in \citealt{giacobbo2018}), and $W_{\rm kick}$ is randomly drawn from the Maxwellian distribution derived by   \cite{hobbs2005}. According to this formalism, if a BH forms by prompt collapse of the parent star $V_{\rm kick}=0$.

If the SN occurs when the BH or NS progenitor is member of a binary, the SN kick can unbind the system. The survival of the binary system depends on the orbital elements at the moment of the explosion and on the SN kick. If the binary remains bound, its post-SN semi-major axis and eccentricity are calculated as described in the appendix A1 of \cite{hurley2002}.

\subsection{Binary evolution} \label{sec:bse}

\subsubsection{Mass transfer}
Mass transfer has been implemented in {\sc SEVN} following the prescriptions described in \cite{hurley2002} with few important updates. {\sc SEVN} considers both wind mass transfer and Roche lobe overflow. Herebelow we give a summary of our implementation, highlighting the differences with respect to {\sc BSE} \citep{hurley2002}, while we refer to the supplementary material, Appendix C, for more details.

The mean accretion rate by stellar winds is calculated from the \cite{bondi1944} formula, following \cite{hurley2002}. Mass transfer by stellar winds is definitely a non-conservative mass transfer process. Thus, we describe also the change of orbital angular momentum, stellar spin and eccentricity following \cite{hurley2002}.

At every time-step we evaluate whether one of the two members of the binary fills its Roche lobe by calculating the Roche lobe as \citep{eggleton1983}
\begin{equation}\label{eq:roche}
R_{L,i}=a\,{}\frac{0.49\,{}q_i^{2/3}}{0.6\,{}q_i^{2/3}+\ln{(1+q_i^{1/3})}},
\end{equation}
where $q_i=M_i/M_j$ with $i=1$, $j=2$ ($i=2$, $j=1$) for the primary (secondary) star. If $R_1\ge{}R_{\rm L,1}$, mass is transferred from the primary to the secondary. We allow for non-conservative mass transfer, which means that the mass lost by the primary at every time step $\Delta{}m_1$ can be larger than the mass accreted by the secondary $\Delta{}m_2$.

If the Roche-lobe filling donor is a neutron star (NS, $k=13$) or a BH ($k=14$), the accretor must be another NS or BH. In this case, the two objects are always merged.

In all the other cases, to decide the amount of mass transferred from the primary $\Delta{}m_1$, we first evaluate the stability of mass transfer using the radius-mass exponents $\zeta{}$ defined by \cite{webbink1985}. If the mass transfer is found to be unstable over a dynamical timescale, the stars are merged (if the donor is a main sequence or an Hertzsprung-gap star) or enter CE (if the donor is in any other evolutionary phase).

If the mass transfer is stable, the mass loss rate of the primary is described as
\begin{equation}\label{eq:stableroche}
\dot{M}_1=3\times{}10^{-6}\,{}{\rm M}_\odot{}\,{}{\rm yr}^{-1}\,{}\left(\frac{M_1}{{\rm M}_\odot}\right)^2\,{}\left[\ln{(R_1/R_{\rm L,1})}\right]^3.
\end{equation}
This is similar to equation~58 of \cite{hurley2002}, but with an important difference: unlike \cite{hurley2002}, we do not need to put any threshold to the dependence on $M_1^2$ to obtain results that are consistent with {\sc BSE}. The term $[\ln{(R_1/R_1)}]^3$ accounts for the fact that mass loss should increase if the Roche lobe is overfilled. If the primary is a degenerate star, $\dot{M}_1$ is increased by a factor $10^3\,{}M_1/\max(R_1/{\rm R}_\odot,10^{-4})$.

Finally, if mass transfer is dynamically stable but unstable over a thermal time-scale, the mass lost by the primary is calculated as the minimum between the result of equation~\ref{eq:stableroche} and the following equation:
\begin{equation}\label{eq:thunstable}
\dot{M}_1=\begin{cases}\frac{M_1}{\tau_{\rm K1}} & \textrm{if } k= 2, 3, 4, 5, 6, 8, 9 \cr   \frac{M_1}{\tau_{\rm D1}} & \textrm{if } k = 0, 1, 7 \cr \end{cases}
\end{equation}
 where $\tau_{\rm K1}$ is the Kelvin-Helmholtz timescale and $\tau{}_{\rm D1}$ is the dynamical timescale of the donor. These timescales are defined as in \cite{hurley2002}.

 In the case of a stable or thermally unstable mass transfer, if the accretor is a non-degenerate star, we assume that the accretion is limited by the thermal timescale of the accretor, as described by \cite{hurley2002}. In particular, the accreted mass $\Delta{}m_2$ is
\begin{equation}\label{eq:thermallimit}
\Delta{}m_2=\min{\left(\alpha{}_\tau\,{}\frac{M_2}{\dot{M}_1\,{}\tau_{\rm K2}},1\right)} \Delta{}m_1,
\end{equation}
where $\Delta{}m_1$ is the mass lost by the donor, $\tau_{\rm K2}$ is the Kelvin-Helmholtz timescale of the accretor and $\alpha{}_\tau{}$ is a dimensionless efficiency parameter ($\alpha_\tau=10$ according to \citealt{hurley2002}). 

This is a crucial difference with respect to other population-synthesis codes (e.g. {\sc startrack}, \citealt{belczynski2008}), which assume that the accreted mass is $\Delta{}m_2=f_a\,{}\Delta{}m_1$, where $0\leq{}f_a\leq{}1$ is a constant efficiency factor, without accounting for the response of the secondary.

With respect to \cite{hurley2002}, we introduce an important difference in the treatment of a Wolf-Rayet (WR, $k=7, 8, 9$) accretor in a stable or a thermally unstable Roche lobe phase: we assume that if the donor has a Hydrogen envelope ($k=0,1,2,3,4,5,6$), the WR does not accrete any Hydrogen. In contrast, \cite{hurley2002} assume that the WR accretes a Hydrogen envelope, becoming a core Helium burning (cHeB) or an AGB star. We make this choice because the winds of the WR are expected to eject a tiny envelope very fast with respect to our time-steps.

If the accretor is a degenerate star, WD ($k=10,11,12$), NS ($k=13$) or BH ($k=14$), the accreted mass is estimated as:
\begin{equation}\label{eq:eddington}
\Delta{}m_2=\min{(\Delta{}m_1,\Delta{}m_{\rm e})},
\end{equation}
where 
\begin{equation}\label{eq:supedd}
\Delta{}m_{\rm e} = 2.08\times{}10^{-3}{\rm M}_\odot\,{}\,{}f_{\rm Edd}\,{}(1.0 + X)^{-1}\,{}\left(\frac{R_2}{{\rm R}_\odot}\right)\,{}\left(\frac{dt}{\rm yr}\right).
\end{equation}
In equation~\ref{eq:supedd}, $X$ is the Hydrogen fraction of the donor star, $R_2$ is the radius of the accretor (for a BH we use the Schwarzschild radius), $dt$ is the time-step in yr, and $f_{\rm Edd}$ is a dimensionless factor indicating whether we allow for super-Eddington accretion (in this paper we assume $f_{\rm Edd}=1$, which corresponds to Eddington limited accretion).

If the accretor is a WD, we also consider the possibility of nova eruptions, following the treatment of \cite{hurley2002}.

If the mass change (of the donor or the accretor) induced by mass transfer is $\Delta{}m>\gamma{}_{\rm m}\,{}M$ (see equation~\ref{eq:chtracktext}), then {\sc SEVN} finds a new track as described in Section~\ref{sec:interp}.

Finally, the variation of orbital angular momentum and stellar spins induced by non-conservative Roche-lobe overflow mass transfer is implemented as in \cite{hurley2002} and summarized in the supplementary material, Appendix C.

\subsubsection{Common Envelope and Stellar Mergers}

In {\sc SEVN}, a common envelope (CE) evolution is the result of (i) a Roche-lobe overflow  unstable on a dynamical timescale, or (ii) a collision at periapsis between two stars\footnote{A collision happens at periapsis when $(R_1+R_2)>(1-e)\,{}a$.}, or (iii) a contact binary, i.e. a binary where both stars fill their Roche lobes ($R_1\ge{}R_{\rm L,1}$ and $R_2\ge{}R_{\rm L,2}$ at the same time).  

{   In these three aforementioned cases, if the donor is a main sequence (MS) or a Hertzsprung-gap (HG) star the two stars are merged directly, without even calculating the CE evolution. In this case, we assume that the binary will not survive CE evolution, because the donor lacks a well-developed core \citep{dominik2012}}. In contrast, if  the donor star has a well-developed core ($k=3,4,5,6,8,9$), the binary enters the routine calculating the CE phase.

During a CE phase, the core of the donor and the accretor are engulfed by the donor's envelope. They begin to spiral in transferring energy to the CE. If the energy released is sufficient to eject the entire envelope the system survives, otherwise the donor coalesces with the accretor. To derive the outcomes of the CE evolution we follow the same formalism as described by \cite{hurley2002}.

This formalism is based on two parameters \citep{webbink1984,kool1992,ivanova2013}:  $\alpha$ is the fraction of the orbital energy released during the spiral-in phase and converted into kinetic energy of the envelope, and $\lambda$ is a structural parameter used to define the binding energy of the envelope.

We can write the initial binding energy of the CE as 
\begin{equation}\label{eq:lambda}
	E_{\rm bind,i} = - \frac{G}{\lambda}\left(\frac{M_1\,{}M_{1,\mathrm{env}}}{R_1} + \frac{M_2\,{}M_{2,\mathrm{env}}}{R_2}\right)~,
\end{equation}
where $M_{1,\mathrm{env}}$ and $M_{2,\mathrm{env}}$ are the initial masses of the envelope of the primary and of the secondary, respectively.

The fraction of orbital energy which goes into kinetic energy of the envelope is
\begin{equation} \label{eq:alpha}
\Delta{}E_{\rm orb}=-\alpha{}\,{}(E_{\rm orb,f}-E_{\rm orb,i})=\alpha{}\,{}\frac{G\,{}M_{\rm c,1}\,{}M_{\rm c,2}}{2}\left(\frac{1}{a_f}-\frac{1}{a_i}\right),
\end{equation}
where $E_{\rm orb,f}$ ($E_{\rm orb,i}$)  is the orbital energy of the binary after (before) the CE phase, $a_f$ ($a_i$) is the semi-major axis after (before) the CE phase, $M_{\rm c,1}$ and $M_{\rm c,2}$ are the masses of the cores of the two stars. If the secondary is a degenerate remnant or a naked core, then $M_{\rm c,2}$ is the total mass of the star.

By imposing that $E_{\rm bind,i}=\Delta{}E_{\rm orb}$, we can derive the final semi-major axis $a_f$  for which the CE is completely ejected. The binary survives and the entire envelope is ejected  if neither core fills its post-CE Roche lobe, estimated from equation~\ref{eq:roche} assuming $a=a_f$, $q_1=M_{\rm c,1}/M_{\rm c,2}$  and $q_2=M_{c,2}/M_{\rm c,1}$. The resulting post-CE binary has masses $M_1=M_{\rm c,1}$, $M_2=M_{\rm c,2}$ and semi-major axis $a_f$. Then, {\sc SEVN} finds a new track for each naked core (unless the accretor is a compact remnant).

In contrast, the two stars are merged if either of their cores fills its post-CE Roche lobe. We estimate the binding energy of the envelope which remains bound to the system as
\begin{equation}
  E_{\rm bind, f}=E_{\rm bind,i}+\alpha{}\,{}\left(\frac{G\,{}M_{\rm c,1}\,{}M_{\rm c,2}}{2\,{}a_{\rm L}}+E_{\rm orb,i}\right), 
\end{equation}
where $a_{\rm L}$ is the semi-major axis for which the larger core fills its post-CE Roche lobe.

The merger product will have core mass $M_{\rm c,3}=M_{\rm c,1}+M_{\rm c,2}$, total mass $M_3$ and radius $R_3$. To estimate the value of $M_3$ and $R_3$, {\sc SEVN} finds a new track with envelope binding energy equal to $E_{\rm bind,f}$ and with core mass $M_{\rm c,3}$, assuming that the envelope binding energy of the merger product is
\begin{equation}
E_{\rm bind, f}=- \frac{G\,{}M_3\,{}(M_3-M_{\rm c,3})}{\lambda{}\,{}R_3}.
 \end{equation}

The spectral type and the other properties of the merger product are thus uniquely determined by the track found by {\sc SEVN} through this search. This procedure is significantly different with respect to the one implemented by \cite{hurley2002}. In {\sc BSE} the final mass $M_3$ is found by assuming a relation between mass and radius ($R\propto{}M^{-x}$) and then by solving the relation between $M_3$ and the other relevant quantities ($M_{\rm c,3}$, $M_1$, $M_2$, $E_{\rm bind,i}$ and $E_{\rm bind,f}$) numerically. With {\sc SEVN} the values of $M_3$ and $R_3$ are determined self-consistently by the search algorithm.

Another substantial upgrade with respect to {\sc BSE} is that {\sc SEVN} does not need to use a ``matrix of stellar types'' as the one reported in Table~2 of \cite{hurley2002}. In fact, to determine the stellar type of the merger product {\sc BSE} reads a matrix where the type of the merger product is given by the combination of the stellar types of the two merged stars. In contrast, {\sc SEVN} does not need any `artificially' defined spectral type, because the spectral type is the natural result of the search algorithm described above. This holds both for colliding unevolved stars (MS and HG stars) and post-CE mergers.

The only exception to the formalism described above is the case in which a star merges with a BH (or a NS) after a CE phase. In the latter case, we assume that the final object remains a BH (or a NS) and that none of the mass of the donor star is accreted by the BH (or NS).

\subsubsection{Tidal Evolution}

We implement the tidal equilibrium model of \cite{hut1981}, which is based on the weak friction approximation and constant time lag model. 
In this model, the misalignment of the tidal bulges with respect to the perturbing potential allows spin-orbit coupling and dissipation of orbital energy.
We evolve semi-major axis, eccentricity and  spin using the secular averaged equations of \cite{hut1981}:
\begin{align}
&\frac{1}{a}\diff{a}{t} = - 6 \left(\frac{k}{T}\right) q(q+1) \left(\frac{R}{a}\right)^8 \frac{1}{(1 - e^2)^{15/2}}\cdot \\ &\quad \quad \quad \cdot\left\{f_1(e^2)-(1-e^2)^{2/3}f_2(e^2)\frac{\Omega_{\rm spin}}{\Omega_{\rm orb}}\right\} \\
&\frac{1}{e}\diff{e}{t} = - 27 \left(\frac{k}{T}\right) q(q+1) \left(\frac{R}{a}\right)^8 \frac{1}{(1 - e^2)^{13/2}} \cdot\\
&\quad \quad \quad \cdot \left\{f_3(e^2)-\frac{11}{18}(1-e^2)^{2/3}f_4(e^2)\frac{\Omega_{\rm spin}}{\Omega_{\rm orb}}\right\} \\
&\diff{\Omega_{\rm spin}}{t} = 3 \left(\frac{k}{T}\right) \frac{q^2}{r^2_{\rm g}} \left(\frac{R}{a}\right)^6 \frac{\Omega_{\rm orb}}{(1 - e^2)^{6}} \cdot\\
&\quad \quad \quad \cdot \left\{f_2(e^2)-(1-e^2)^{2/3}f_5(e^2)\frac{\Omega_{\rm spin}}{\Omega_{\rm orb}}\right\} \\
\end{align}
where $q$ is the mass ratio between the perturbing star and the star undergoing tides, while $r^2_{\rm g} = I/M\,{}R^2$, $\Omega_{\rm spin}$ and $R$ are the gyration radius, spin and radius of the star undergoing tides, respectively. The $f_i(e^2)$ terms are polynomial functions of the eccentricity given by \cite{hut1981}. In the present work, we assume that the stars have zero obliquity, i.e. the spin is aligned with the angular momentum vector of the binary.

The term $k/T$ determines the timescale of the tidal evolution and depends on the dissipation mechanism responsible for the misalignment of the tidal bulges. We adopt the prescriptions of \cite{hurley2002}, which are based on \cite{zahn1975} for the tide in radiative envelopes and \cite{zahn1977} for the tide in convective envelopes (see also \citealt{rasio1996}).

\subsubsection{Gravitational-wave Decay}
Gravitational-wave (GW) decay is implemented in {\sc SEVN} according to the formulas by \cite{peters1964}, which describe the loss of energy and angular momentum of a system due to the radiation of GWs. In particular, the loss of orbital angular momentum and the loss of eccentricity due to GW emission are estimated as
\begin{eqnarray}
\frac{\dot{J}_{\rm orb}}{J_{\rm orb}}=-\frac{32}{5}\,{}\frac{G^3}{c^5}\,{}\frac{M_1\,{}M_2\,{}(M_1+M_2)}{a^4}\,{}\frac{1+\frac{7}{8}\,{}e^2}{(1-e^2)^{5/2}}\label{eq:peters1}\\ 
\frac{\dot{e}}{e}=-\frac{32}{5}\,{}\frac{G^3}{c^5}\,{}\frac{M_1\,{}M_2\,{}(M_1+M_2)}{a^4}\,{}\frac{\frac{19}{9}+\frac{121}{96}\,{}e^2}{(1-e^2)^{5/2}}\label{eq:peters2}
\end{eqnarray}
Equations ~\ref{eq:peters1} and \ref{eq:peters2} are evaluated for all double compact-object binaries ($k\ge{}10$) and not only for the closest ones (in contrast, {\sc BSE} calculates the GW decay only if $a\leq{}10 R_{\odot}$).

\subsection{Comparison of {\sc SEVN} with {\sc BSE} and {\sc MOBSE}} 

\begin{figure*}
	\includegraphics[width=\hsize]{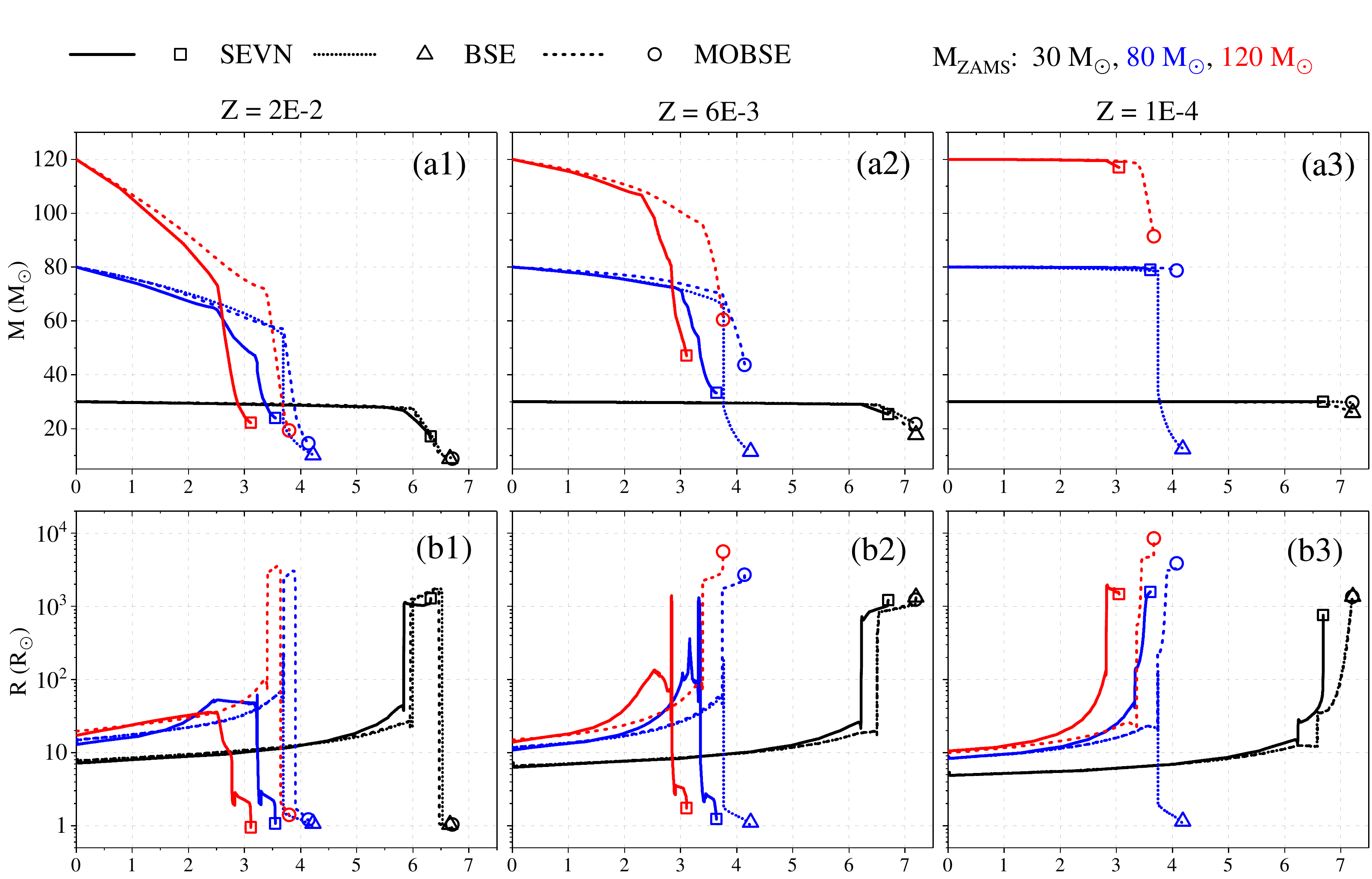}
	\includegraphics[width=\hsize]{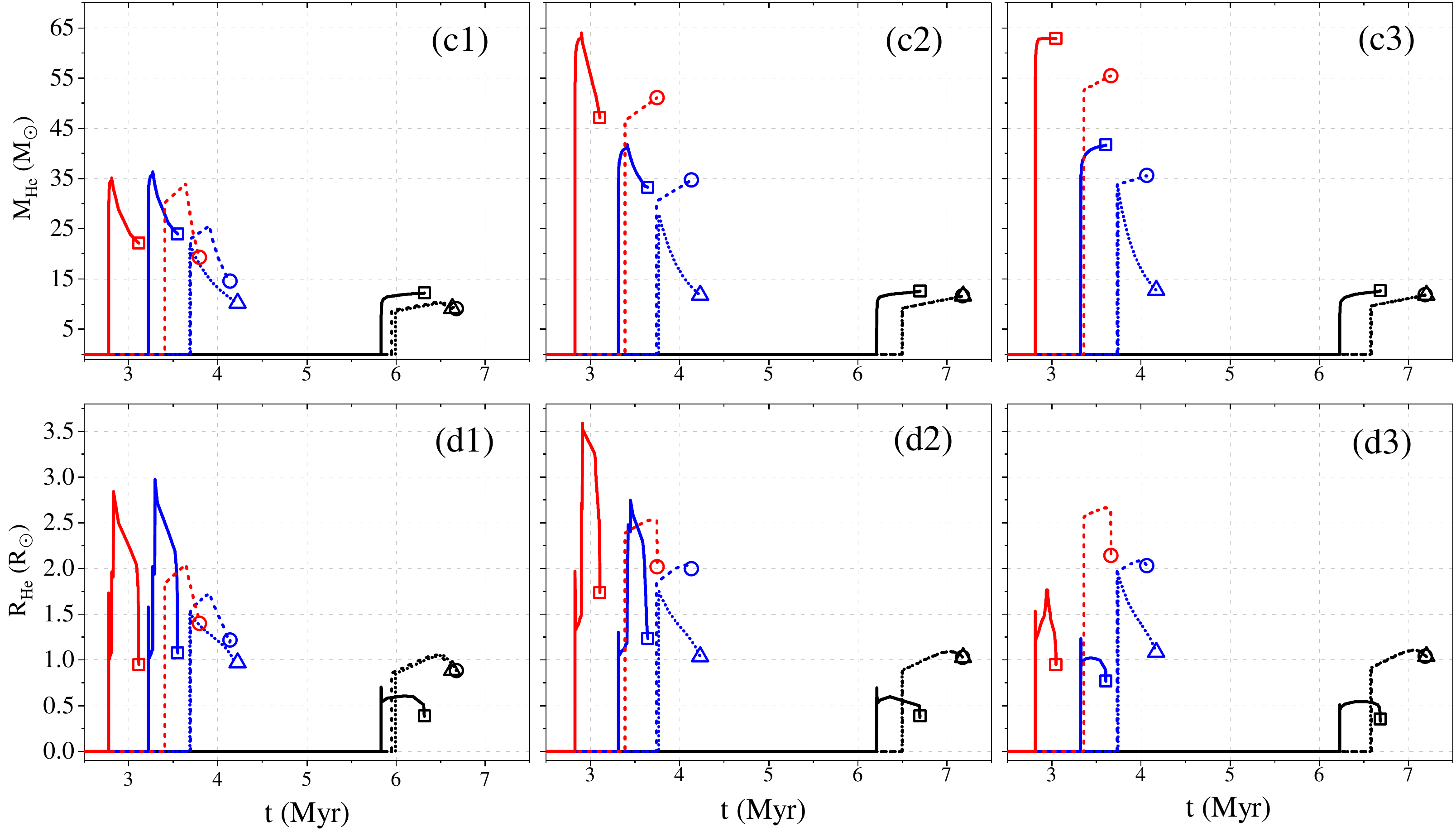}
	\caption{Time evolution of the physical stellar parameters of different stars, derived with {\sc BSE}, {\sc MOBSE} and {\sc SEVN}. The black, blue and red lines refer to a star with $M_{\mathrm{ZAMS}}=30\msun{}$, $80\msun{}$ and  $120\msun{}$, respectively. Top row (i.e. panels labelled with \textit{a}): total stellar mass; second row (\textit{b}): stellar radius; third row (\textit{c}): He-core mass; bottom row (\textit{d}): He-core radius. Left-hand column (i.e. panels labelled with \textit{1}): metallicity $Z=0.02$; central column (\textit{2}):  $Z=6\times 10^{-3}$; right-hand column (\textit{3}): $Z=10^{-4}$. Solid lines: \sevn{}; dotted lines: \bse{}; dashed lines: \mobse{}. The open squares identify the final point of the curves obtained with \sevn{} (open triangles: \bse{}; open circles: \mobse{}). We do not evolve the star with $M_\mathrm{ZAMS} = 120\msun{}$ with \bse{}, because the fitting formulas included in BSE might be inaccurate for $M_{\mathrm{ZAMS}} > 100\msun{}$.}
	\label{fig:comparison}
\end{figure*}

Fig. \ref{fig:comparison} shows the time evolution of the total star mass ($M$), the stellar radius $R$, the He core mass $M_{He}$, and the He radius $R_{He}$ for three selected massive stars ($M_{\mathrm{ZAMS}}=30, 80,$ and $120 \msun{}$), at different metallicity $Z$. The results obtained with the \sevn{} code (solid lines) are compared to those obtained with \bse{} (dotted lines) and \mobse{} (dashed lines). The star with $M_\mathrm{ZAMS}=120\msun{}$ is not evolved with the \bse{} code because the fitting formulas implemented in \bse{} may be inaccurate for $M_\mathrm{ZAMS}>100\msun{}$ \citep{hurley2002}.

From Fig. \ref{fig:comparison} it is apparent that the star lifetime  in \sevn{} is up to $\sim 30\%$ shorter than that obtained with \mobse{} and \bse{}. 
\sevn{} and \mobse{} show a similar evolution of $M$ for all considered metallicities $Z$ and for all selected $M_\mathrm{ZAMS}$ (Fig. \ref{fig:comparison}, panels \textit{a1}, \textit{a2} and \textit{a3}). In contrast, \bse{} predicts a different evolution for the $80\msun{}$ star, especially in the late evolutionary stages for $Z \lesssim 6\times 10^{-3}$, because of different stellar wind models. The difference is maximum at $Z=10^{-4}$ (panel \textit{a3}) where \bse{} predicts the formation of a WR star with $M\simeq 12\msun{}$ while \sevn{} forms a red hypergiant star with $M\simeq 80\msun{}$.

The evolution of $R$ shows even more differences. According to \sevn{}, at $Z=2\times 10^{-2}$ (panel \textit{b1}), the stars with $M_\mathrm{ZAMS}=80$ and $120 \msun{}$ become WR stars before reaching the red giant branch, therefore their radius is always $< 80 \rsun{}$. In contrast, in \mobse{} they become WR stars at a later stage, after having already gone through the red giant branch and having reached $R > 2\times 10^3 \rsun{}$.

Furthermore, for the $30\,{}\msun{}$ star, both \bse{} and \mobse{} predict the formation of a WR star ($R=R_{He}\simeq 1\rsun{}$) while, in \sevn{}, the star ends its life as a red supergiant ($R\simeq10^3 \rsun{}$) with a Hydrogen-envelope mass of $\sim 5\msun{}$.

According to \sevn{}, at $Z=6\times 10^{-3}$ (panel \textit{b2}), both the $80\msun{}$ and  the $120\msun{}$ star die as WR stars, with $R\lesssim 2\rsun{}$. In contrast, according to \mobse{}, the same stars die as red supergiants with $R \gtrsim 3\times 10^3 \rsun{}$.

The evolution of $R$ is quite similar at $Z=10^{-4}$ (panel \textit{b3}), even though \sevn{} forms stars with smaller radii compared to those formed with \mobse{} ($\sim{}1.5\times 10^{3}\rsun{}$ against $\gtrsim 4\times 10^{3}\rsun{}$).

The three codes show a quite similar evolution of $M_{He}$ (panels \textit{c1}, \textit{c2} and \textit{c3}), with \sevn{} forming slightly more massive He cores (up to $15\%$) at $Z=10^{-4}$.

Furthermore, \sevn{} forms He-core radii up to $70\%$ smaller than those obtained with \mobse{} and \bse{},  except for the $80\msun{}$ and the $120\msun{}$ stars at $Z=2\times 10^{-2}$ (panels \textit{d1}, \textit{d2} and \textit{d3}).

\subsection{Initial conditions}

We have run 15 sets of simulations with metallicity $Z = 4\times 10^{-2}, 3\times 10^{-2}, 2\times 10^{-2}, 1.6\times 10^{-2}, 10^{-2}, 8\times 10^{-3}, 6\times 10^{-3}, 4\times 10^{-3}, 2\times 10^{-3}, 1.6\times 10^{-3}, 10^{-3}, 8\times 10^{-4}, 4\times 10^{-4}, 2\times 10^{-4}, 10^{-4}$, respectively. Each simulation set consists of $10^7$ binary systems. We used the same set of initial conditions for all simulations. The masses of the primary stars ($M_1$) are drawn from a Kroupa initial mass function  (IMF, \citealt{kroupa2001})

\begin{equation}
\xi\left(M_1\right)\propto M_1^{-2.3}\,\,\,\,\,\,\,\,\,\,M_1\in\left[10,150\right]\msun{}.
\end{equation}

We chose $10\msun{}$ as the lower mass limit of the IMF because in this paper we focus only on the formation and evolution of BH binaries. We will extend the IMF range in forthcoming works.

The masses of the secondary stars ($M_2$) are distributed according to \citet{sana2012}

\begin{equation}
\xi\left(q\right)\propto q^{-0.1}\,\,\,\,\,\,\,\, q=\frac{M_2}{M_1}\in\left[0.1, 1\right]\,\,\,\,\,\, \mathrm{and}\,\,\, M_2 \geq 10\msun{}.
\end{equation}

The initial orbital periods ($\mathcal{P}$) and eccentricities ($e$) also follow the distributions given by \citet{sana2012},
\begin{equation}
\xi\left(\mathcal{P}\right)\propto \mathcal{P}^{-0.55}\,\,\,\,\,\,\,\, \mathcal{P}=\log\left(P/\mathrm{day}\right)\in\left[0.15, 5.5\right],
\end{equation}
\begin{equation}
\xi\left(e\right)\propto e^{-0.42}\,\,\,\,\,\,e\in\left[0,1\right].
\end{equation}

We evolve each binary system for $20$ Myr to ensure that both stars have ended their evolution by the end of the simulation. Furthermore, we adopt the \textit{rapid} model for all the SN explosions {   and $\left(\alpha, \lambda\right) = \left(1, 0.1\right)$ for the common envelope phase.}

\section{Results}
\label{sec:results}

        
        \begin{figure*}
        	\includegraphics[width=\hsize]{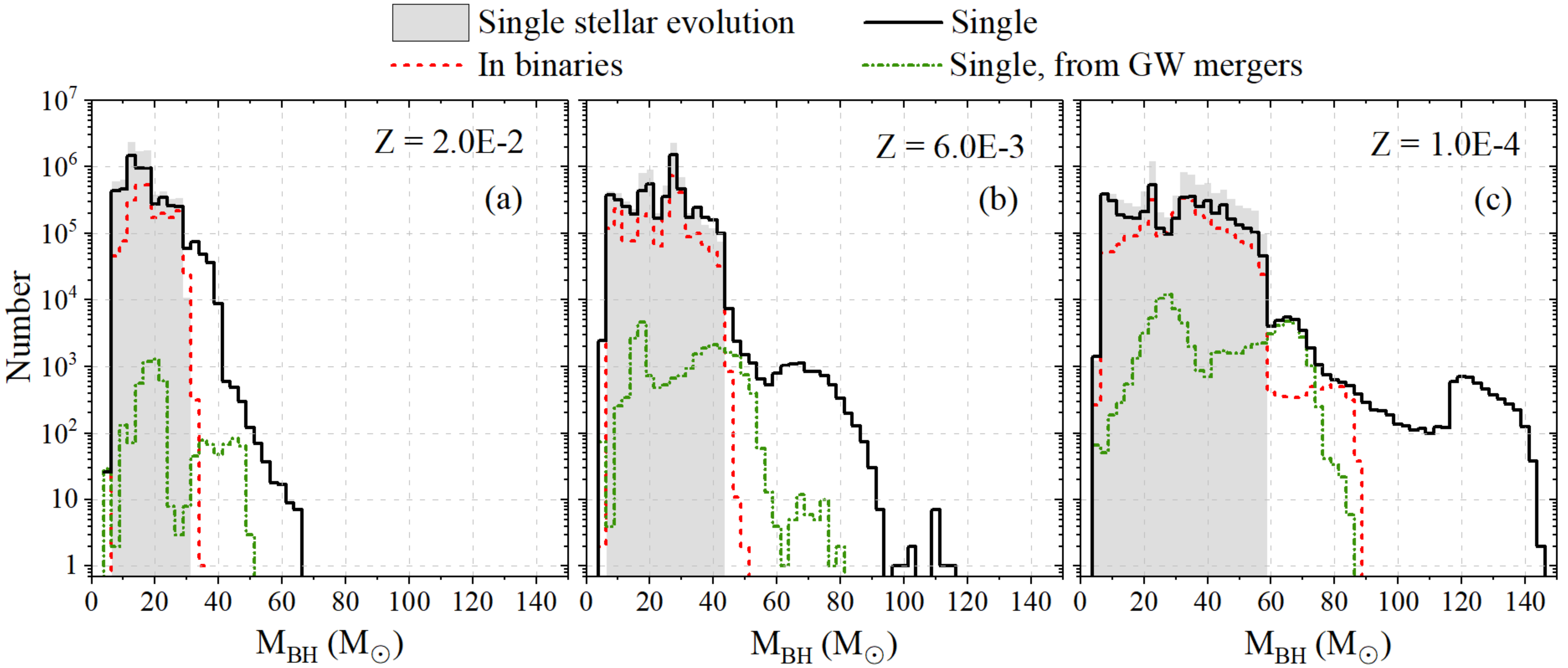}
        	\caption{Distribution of the masses of BHs formed in our simulations. Dashed red line: BHs in compact-object binaries; solid black line: single BHs; dash-dotted green line: single BHs that formed from a GW merger. Grey area: BH mass distribution from single star evolution. Left-hand panel (\textit{a}): $Z=2\times 10^{-2}$; central panel (\textit{b}): $Z=6\times 10^{-3}$; right-hand panel (\textit{c}): $Z=10^{-4}$. }
        	\label{fig:nbh}
        \end{figure*}

        Figure \ref{fig:nbh} shows the distribution of  BH masses in our simulations, at different metallicity. We show the masses of single BHs (solid black line), single BHs that form from GW mergers (dash-dotted green line) and BHs which are members of compact-object binaries (dashed red lines). We stress that all BHs at the end of our simulations are either single or members of compact-object binaries, because all stars have turned to compact objects by the end of the simulations.
        
        Fig. \ref{fig:nbh} also shows that the mass distribution of BHs in compact-object binaries is not significantly different from the one we obtain from single stellar evolution (grey area in Fig. \ref{fig:nbh}). 
        
        In contrast, the distribution of masses of single BHs is very different, especially at low $Z$. At $Z=6\times 10^{-3}$ ($Z=10^{-4}$) we form single BHs with mass up to $\sim 90\msun{}$ ($145 \msun{}$), while the maximum mass of BHs in compact-object binaries is $\sim 40\msun{}$ ($\sim{}90\msun{}$). Most massive single BHs come from the merger of an evolved star with a MS star, and only a small fraction of them come from GW mergers (see Sec. \ref{sec:discussion} for a detailed discussion).

\begin{figure*}
	\includegraphics[width=\hsize]{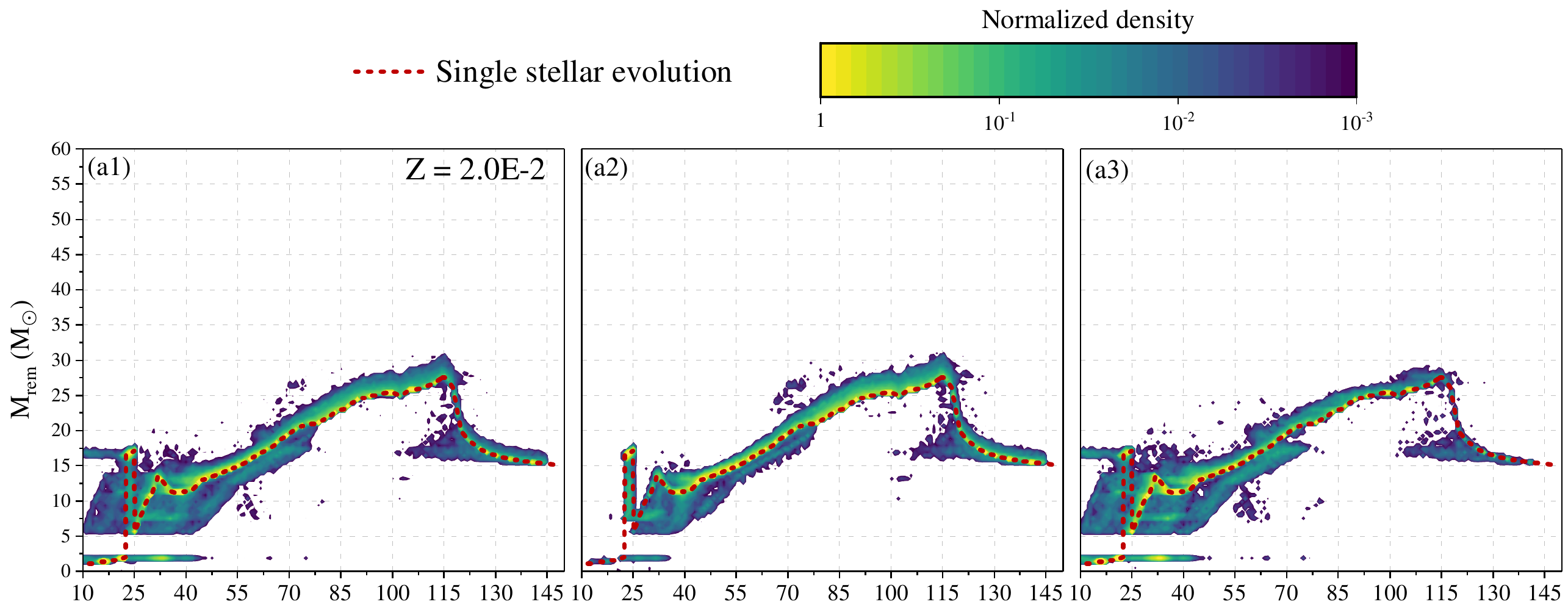}
	\includegraphics[width=\hsize]{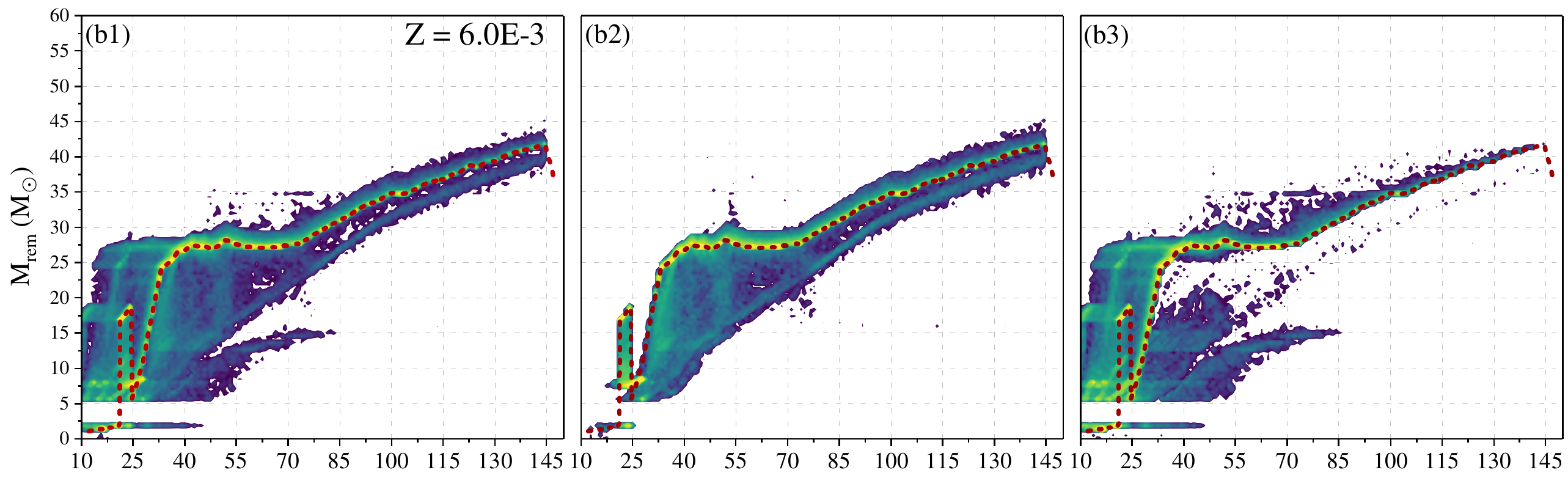}
	\includegraphics[width=\hsize]{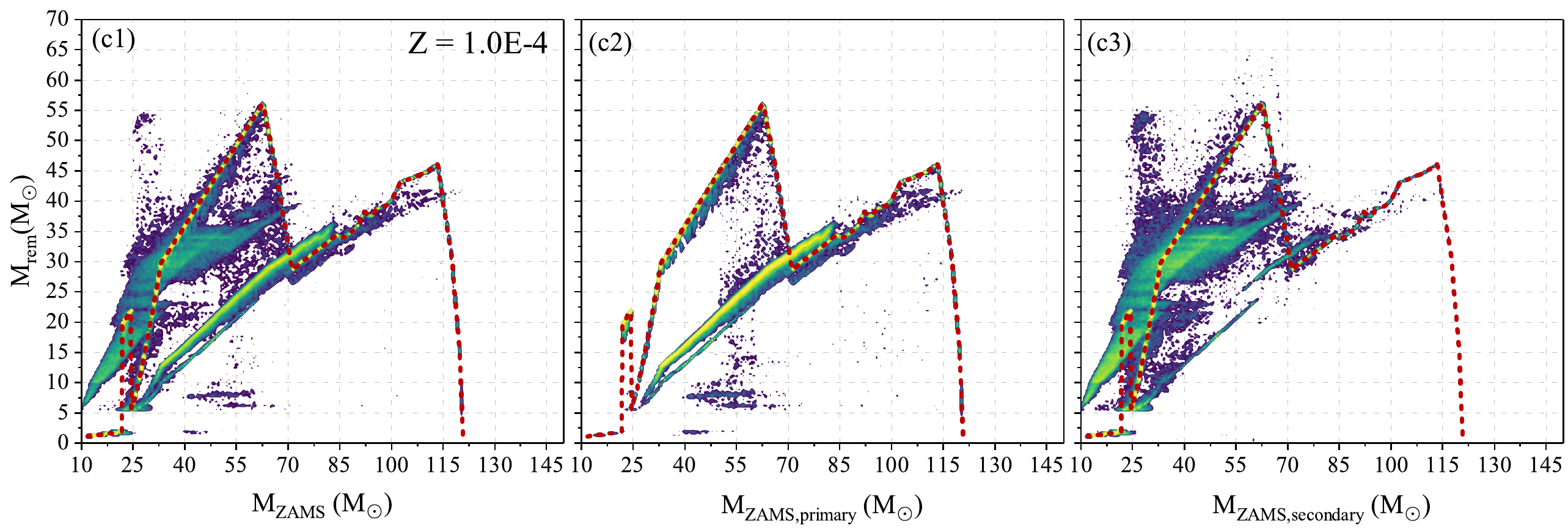}
	\caption{Mass of compact remnants ending up in compact-object binaries, as a function of the ZAMS mass of the progenitor star. The logarithmic colour bar represents the number of compact objects per cell, normalized to the maximum cell-value of each plot. Each cell is a square with a side of $0.5\msun{}$. Rows labelled as \textit{a}, \textit{b} and \textit{c} show the mass spectrum of compact remnants at metallicity $Z=2\times 10^{-2}$, $Z=6\times 10^{-3}$ and $Z=10^{-4}$, respectively. Columns labelled as \textit{1} show all compact remnants; columns labelled as \textit{2} (\textit{3}) show only the compact remnants formed from the primary (secondary) star. The dashed line is the mass spectrum of compact objects obtained from single stellar evolution calculations.}
	\label{fig:spectrum002}
\end{figure*}

Figure \ref{fig:spectrum002} shows the mass spectrum of compact remnants which are members of double compact-object binaries. 
The first column (\textit{a}) shows all compact objects, while the second (\textit{b}) and the third column (\textit{c}) show only the compact objects which form from the primary and the secondary star\footnote{For primary and secondary star we mean the more massive and the less massive member of the binary in the ZAMS.}, respectively.

At $Z=2\times 10^{-2}$, the BHs in compact-object binaries have masses in the range $\left[5, 30\right]\msun{}$, with the heaviest BHs formed from stars with $M_\mathrm{ZAMS}\simeq 115\msun{}$. BHs can be more massive at low metallicity because their progenitor stars lose less mass through stellar winds during their life. The most massive BHs at $Z=6\times 10^{-3}$ have mass $\sim 45\msun{}$ and they form from stars with $M_\mathrm{ZAMS}\simeq 145\msun{}$.

At $Z=10^{-4}$, the heaviest BHs ($\sim 55\msun{}$) form from stars with $M_\mathrm{ZAMS}\simeq 62\msun{}$, that is they do not form from the collapse of the most massive stars. This happens because PPISNe significantly enhance the mass loss of stars with $60\leq M_\mathrm{ZAMS}/\msun{}\leq 115$ and PISNe cause the disintegration of the stars with $M_\mathrm{ZAMS}\gtrsim 120\msun{}$.

From Fig. \ref{fig:spectrum002} it is also apparent that most compact remnants distribute along the curve obtained from single stellar evolution calculations (dashed line). These remnants come from binary stars that evolved through no (or minor) mass transfer episodes. 

In contrast, primary stars that underwent a Roche-lobe overflow episode, or that have lost their envelope after a CE phase, tend to form smaller compact objects than they would have formed if they were evolved as single stars. This is apparent in panels \textit{a2}, \textit{b2}, and \textit{c2} of Fig. \ref{fig:spectrum002}, where compact objects formed by primary stars tend to fall below the single stellar evolution curve. 

Panel \textit{c2} is a particularly significant case: most primaries with ZAMS mass between $\sim{}25$ and $\sim{}85$ M$_\odot$ at $Z=10^{-4}$ become compact remnants with a factor of $\sim{}2-3$ lower mass than compact remnants born from single stars with the same ZAMS mass. These primary stars undergo Roche lobe overflow followed by CE evolution and are completely stripped of their Hydrogen envelope, becoming WR stars.

The secondary stars with $M_\mathrm{ZAMS}\lesssim 30\msun{}$ that accreted mass from the primary star tend to form more massive compact objects than they would have formed if they were single stars (see panels \textit{a3}, \textit{b3}, and \textit{c3}).

More massive secondaries ($M_\mathrm{ZAMS}\gtrsim 30\msun{}$) either fill their Roche lobe at later stages, after they have become more massive than the primary (which has typically evolved into a compact object) or undergo CE; so they lose significant mass and form compact remnants that fall below the single stellar evolution curve (e.g. panel \textit{b3} for $25\leq M_\mathrm{ZAMS}/\msun{}\leq 85$ and $5\leq M_\mathrm{rem}/\msun{}\leq 20$).

The deviation from the mass spectrum obtained from single stellar evolution is more pronounced at low metallicity. This happens because stellar winds are quenched at low metallicity, therefore the mass that can be exchanged during a Roche lobe overflow episode or lost during a CE phase is significantly larger at low $Z$.

It is also worth noting that the mass range of BHs in compact-object binaries is very similar to that obtained from single stellar evolution calculations. In particular, it is very unlikely to find BHs in binaries with a mass significantly larger than the maximum BH mass obtained from single-star evolution, for every metallicity.

From Fig. \ref{fig:spectrum002} it is also apparent that we have a mass gap between the heaviest NS ($\sim 2\msun{}$) and the lightest BH ($\sim 5\msun{}$). This is a feature of the adopted \textit{rapid} SN explosion model that reproduces the observed mass gap between NS and BH masses.

\begin{figure*}
	\includegraphics[width=\hsize]{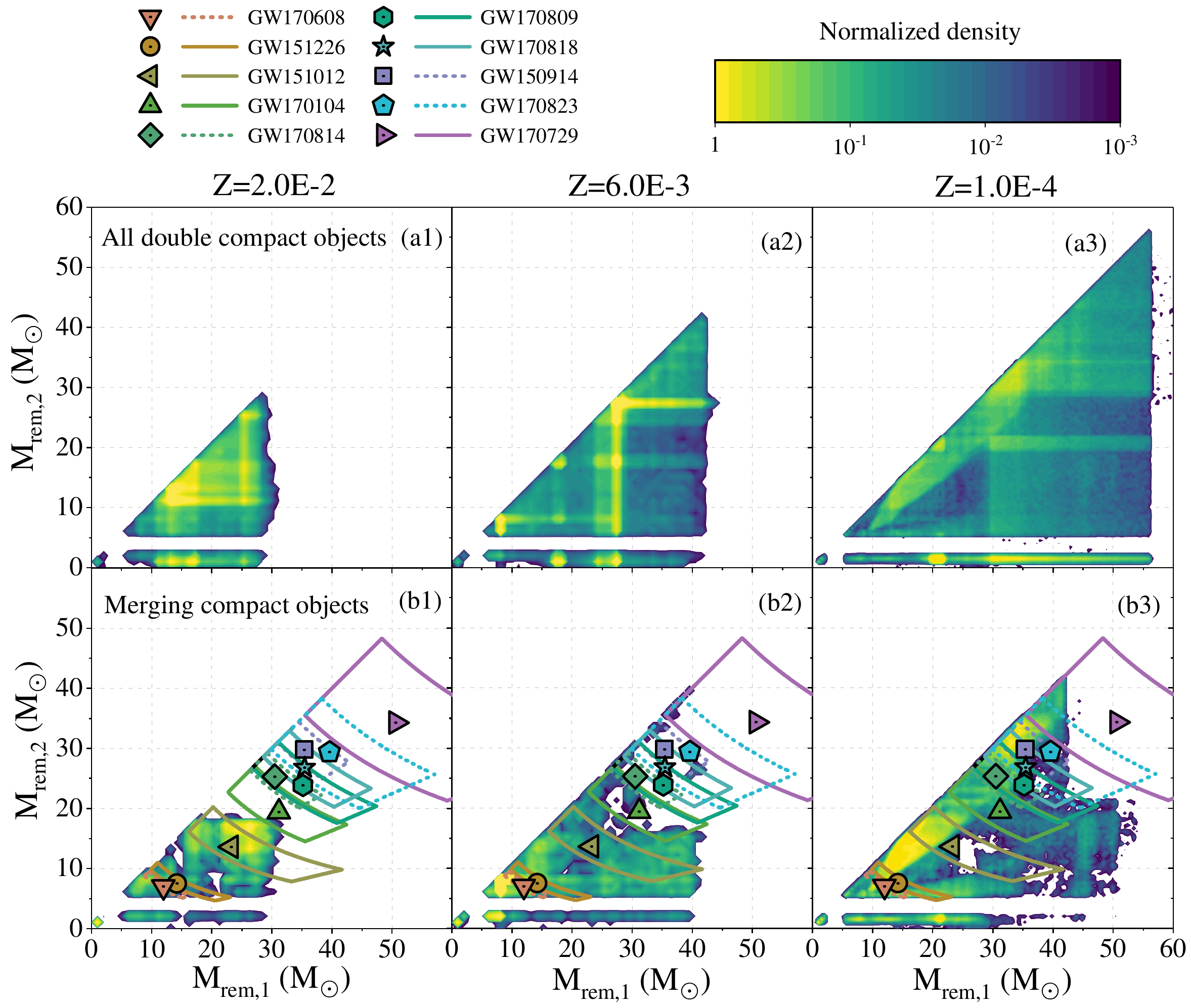}
	\caption{Mass of the less massive remnant (M$_{\rm rem,\,{}2}$) as a function of the mass of the more massive remnant (M$_{\rm rem,\,{}1}$) in all compact-object binaries (top row, labelled as $a$), and in the compact-object binaries merging within a Hubble time  (bottom row, labelled as $b$). The logarithmic colour bar represents the number of remnants per cell, normalized to the maximum cell-value of each plot. Each cell is a square with a side of $0.5\msun{}$. Left-hand column (labelled as \textit{1}): $Z=2\times 10^{-2}$; central column (\textit{2}: $Z=6\times 10^{-3}$); right hand column (\textit{3}): $Z=10^{-4}$. {   The symbols are the BH mergers detected by LIGO/Virgo in O1 and O2. The solid and dashed lines around the symbols define the 90\% credible interval on the chirp mass and the mass ratio of each GW event.} A version of this figure containing all the other considered metallicities is shown in the supplementary material, Appendix D.}
	\label{fig:cobm1m2}
\end{figure*}

Figure \ref{fig:cobm1m2} shows the mass of the less massive remnant as a function of the mass of the more massive remnant, for all double compact-object binaries (\textit{a} panels, in the top row) and for all compact-object binaries merging  within a Hubble time (\textit{b} panels, in the bottom row). 

In the panels of the top row of Fig.~\ref{fig:cobm1m2}  we find a large number of BHs in the areas where the mass spectrum of compact remnants from single stars (see the dashed line of Fig.~\ref{fig:spectrum002}) is quite flat. For instance, at $Z=2\times 10^{-2}$ all stars with $30\leq M_\mathrm{ZAMS}/\msun{}\leq 65$ form BHs with masses between $10\msun{}$ and $17\msun{}$ (cf. panel \textit{a1} of Fig. \ref{fig:spectrum002}).

Fig. \ref{fig:cobm1m2} shows that merging BHs with masses consistent with {   GW151226, GW170608 and GW151012 (i.e. the low-mass GW events)} form at all metallicities in our simulations. {   GW150914, GW170104, GW170809, GW170814, GW170818 and GW170823 (i.e. the GW events hosting BHs with $M_{\rm BH}\ge{}30$ M$_\odot$)} are perfectly matched by the masses of simulated merging BHs at low metallicity ($Z=10^{-4}$, panel \textit{b3}), while merging BHs with mass  $>20\msun{}$ do not form in our simulations at $Z=2\times 10^{-2}$. The 90\% credible levels for the masses of {   GW150914, GW170104, GW170809, GW170814, GW170818 and GW170823} partially overlap with our simulated merging BHs at $Z=6\times 10^{-3}$ (see panels \textit{b1} and \textit{b2}, respectively). {   From Fig. \ref{fig:cobm1m2} it is also apparent that it is unlikely to find merging BHs with masses consistent with GW170729 (i.e. the GW event with the heaviest BHs). The 90\% credible levels for the masses of GW170729 partially overlap with our merging BHs only at $Z=10^{-4}$. On the other hand, dynamical processes might easily lead to the formation of GW170729-like systems \citep{dicarlo2019}.}

It is also worth noting that the most massive BHs formed in our simulations are unlikely to merge within a Hubble time via GWs. Specifically, at $Z=2\times 10^{-2}$ we do form compact-object binaries with both BHs more massive than $\sim 20\msun{}$ (upper-triangular area of panel \textit{a1}) but they do not merge via GWs (the same triangular area is missing in panel \textit{b1}). We obtain the same result at $Z=6\times 10^{-3}$ for BHs with mass $\gtrsim 25\msun{}$ and at $Z=10^{-4}$ for BHs with mass $\gtrsim 40\msun{}$.


\begin{figure}
	\includegraphics[width=\hsize]{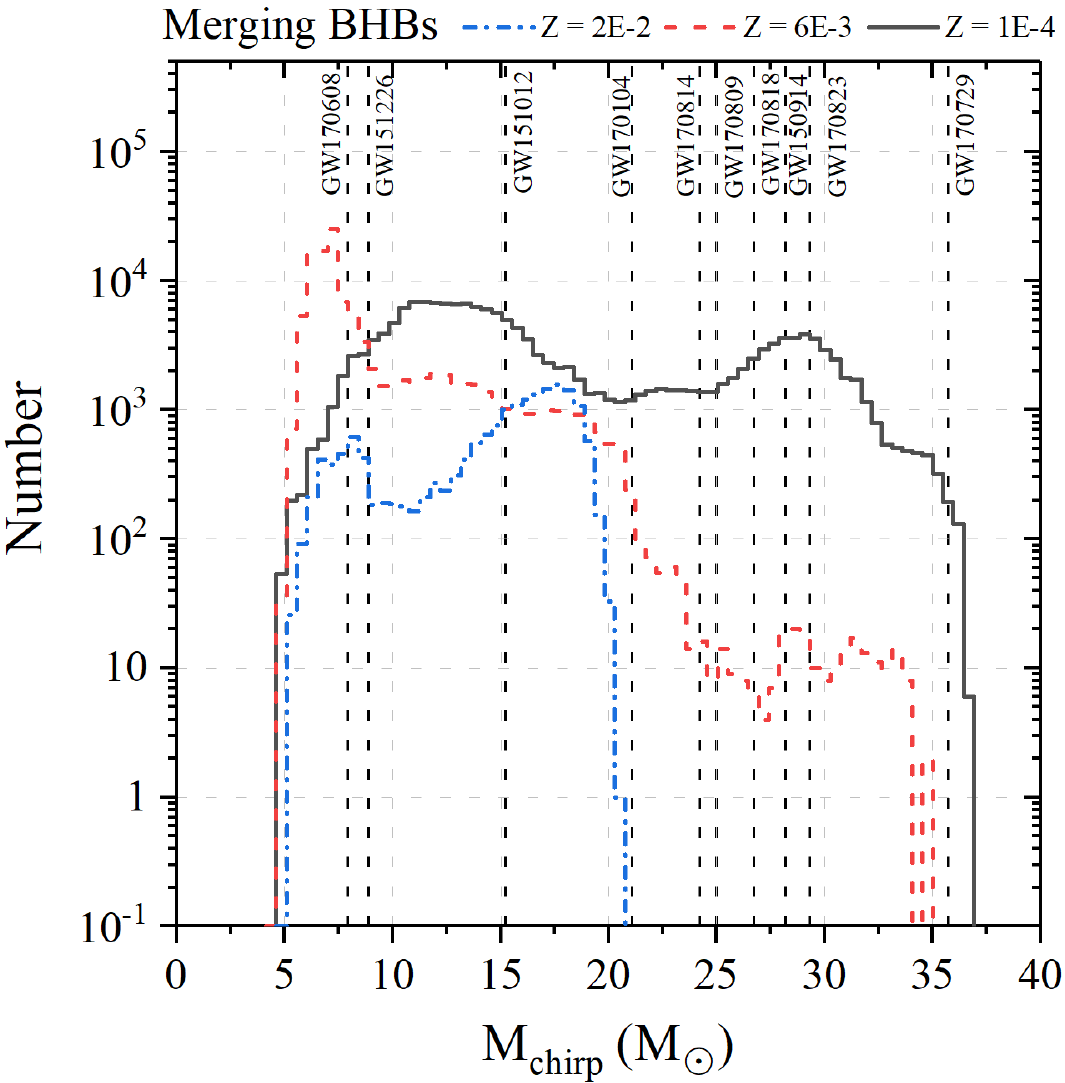}
	\caption{Chirp mass distributions of merging BHBs. The different lines show the results at different metallicities. Dash-dotted blue line: $Z=2\times 10^{-2}$; dashed red line: $Z=6\times 10^{-3}$; solid black line: $Z=10^{-4}$. {   The vertical dashed lines represent the chirp masses of the GW detections. From left to right: GW170608, GW151226, GW151012, GW170104, GW170814, GW170809, GW170818, GW150914, GW170823, and GW170729.} A version of this figure containing all the other considered metallicities is shown in the supplementary material, Appendix D.}
	\label{fig:chirp}
\end{figure}

Figure \ref{fig:chirp} shows the distribution of the chirp masses of merging BHBs at different metallicity. It is apparent that merging BHBs with $M_\mathrm{chirp}\gtrsim 20\msun{}$ (such as GW150914 and GW170814) cannot form at $Z=2\times 10^{-2}$. Merging BHBs with $20\leq M_\mathrm{chirp}\leq 35$ are also unlikely to form at $Z=6\times 10^{-3}$, while they are quite common at $Z\lesssim 10^{-4}$. Furthermore, in our simulations we do not find merging BHs with $M_\mathrm{chirp}\gtrsim 37\msun{}$, independently of metallicity.

\begin{figure}
	\includegraphics[width=\hsize]{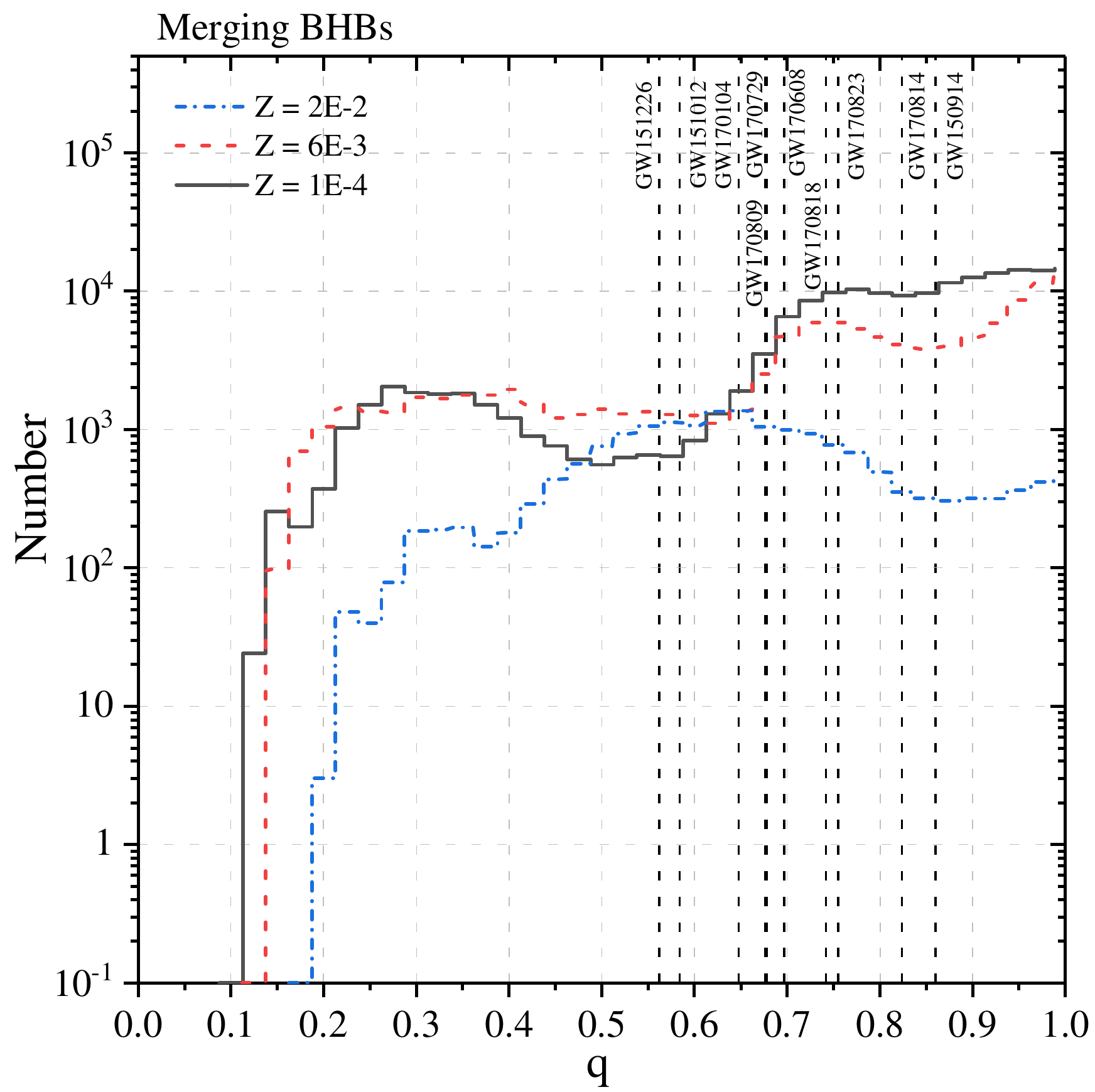}
	\caption{Distribution of the mass ratio of merging BHBs. The different lines show the results at different metallicities. Dash-dotted blue line: $Z=2\times 10^{-2}$; dashed red line: $Z=6\times 10^{-3}$; solid black line: $Z=10^{-4}$. The vertical dashed lines represent the mass ratio the GW detections. From left to right: {   GW151226, GW151012, GW170104, GW170729, GW170809, GW170608, GW170818, GW170823, GW170814, and GW150914.} A version of this figure containing all the other considered metallicities is shown in the supplementary material, Appendix D. }
	\label{fig:ratio}
\end{figure}

Figure \ref{fig:ratio} shows the distribution of the mass ratio ($q=\frac{M_2}{M_1}\,\,,\,\,M_1\geq M_2$) of merging BHBs. At all metallicities, most merging BHs have $q>0.5$, but the fraction of systems with lower mass ratio is not negligible, especially at low $Z$ where the merging BHs with $q<0.5$ are $\sim 10\%$ of the total. We do not find merging BHs with $q<0.1$, and very low mass ratios ($0.1<q<0.2$) seem to be possible only at low metallicity. From Fig. \ref{fig:ratio} it is also apparent that we match the mass ratios of GW detections at all the considered metallicities.

\begin{figure*}
	
	\includegraphics[width=\hsize]{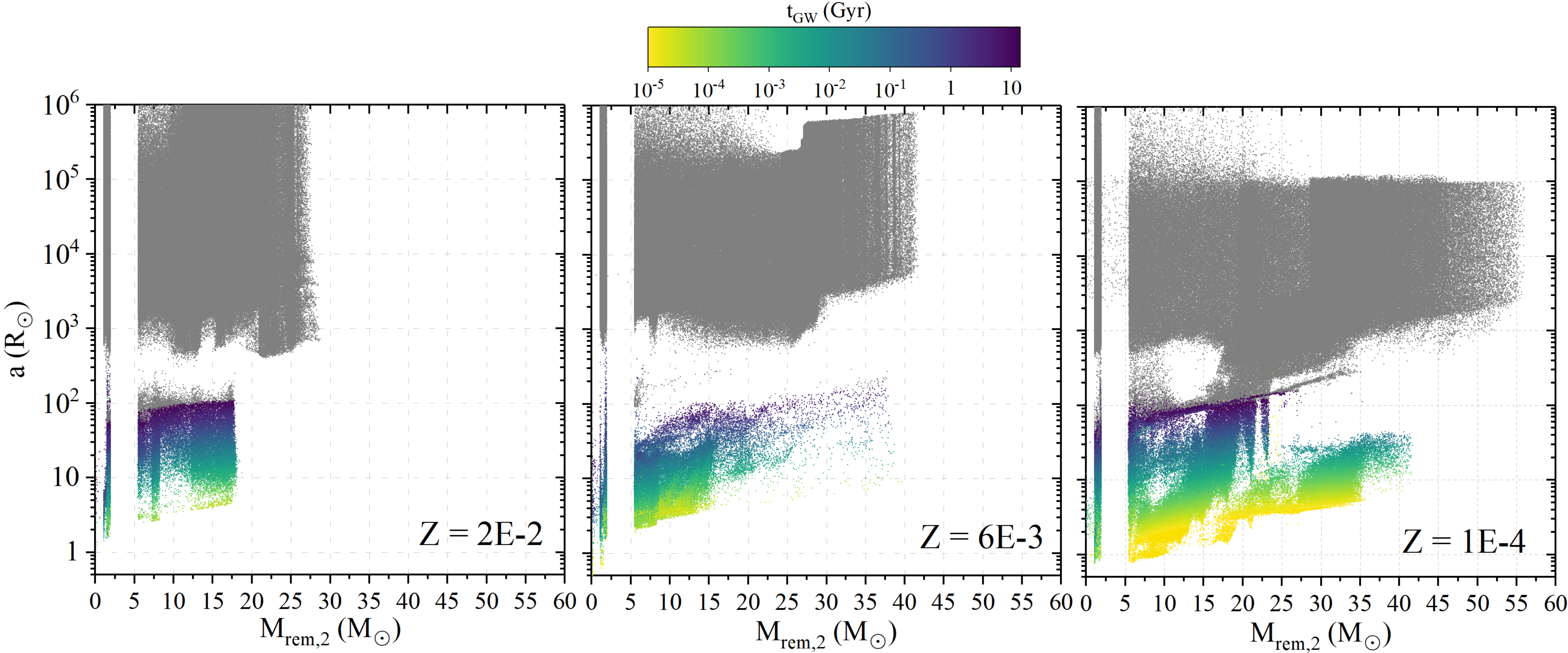}
	\caption{Semi-major axis as a function of the mass of the less massive remnant in double compact-object binaries. The semi-major axis is taken at the time of the formation of the second compact remnant. Left-hand panel \textit{(a)}: $Z=2\times 10^{-2}$; central panel \textit{(b)}: $Z=6\times 10^{-3}$; right-hand panel \textit{(c)}: $Z=10^{-4}$. The logarithmic colour bar represents the GW merger time-scale \citep{peters1964}. Gray points show the systems that do not merge within a Hubble time.}
	\label{fig:a_vs_m}
\end{figure*}

\section{Discussion}
\label{sec:discussion}

\subsection{Dearth of massive BHB mergers}
We have shown in Section \ref{sec:results} that the BHBs with the heaviest BH members are unlikely to merge within a Hubble time (see Fig. \ref{fig:cobm1m2}). This happens because the separation of two massive BHs at the time of the formation of the second remnant is generally too large to let the BHs merge via GWs. This is apparent from Figure \ref{fig:a_vs_m}, which shows the semi-major axis of double compact objects at the time of the formation of the second remnant as a function of the mass of the less massive compact object. At all metallicities, most of the heaviest BHBs have quite large semi-major axes ($a > 10^2 \rsun{}$, grey points).

\begin{figure*}
	\includegraphics[width=0.95\hsize]{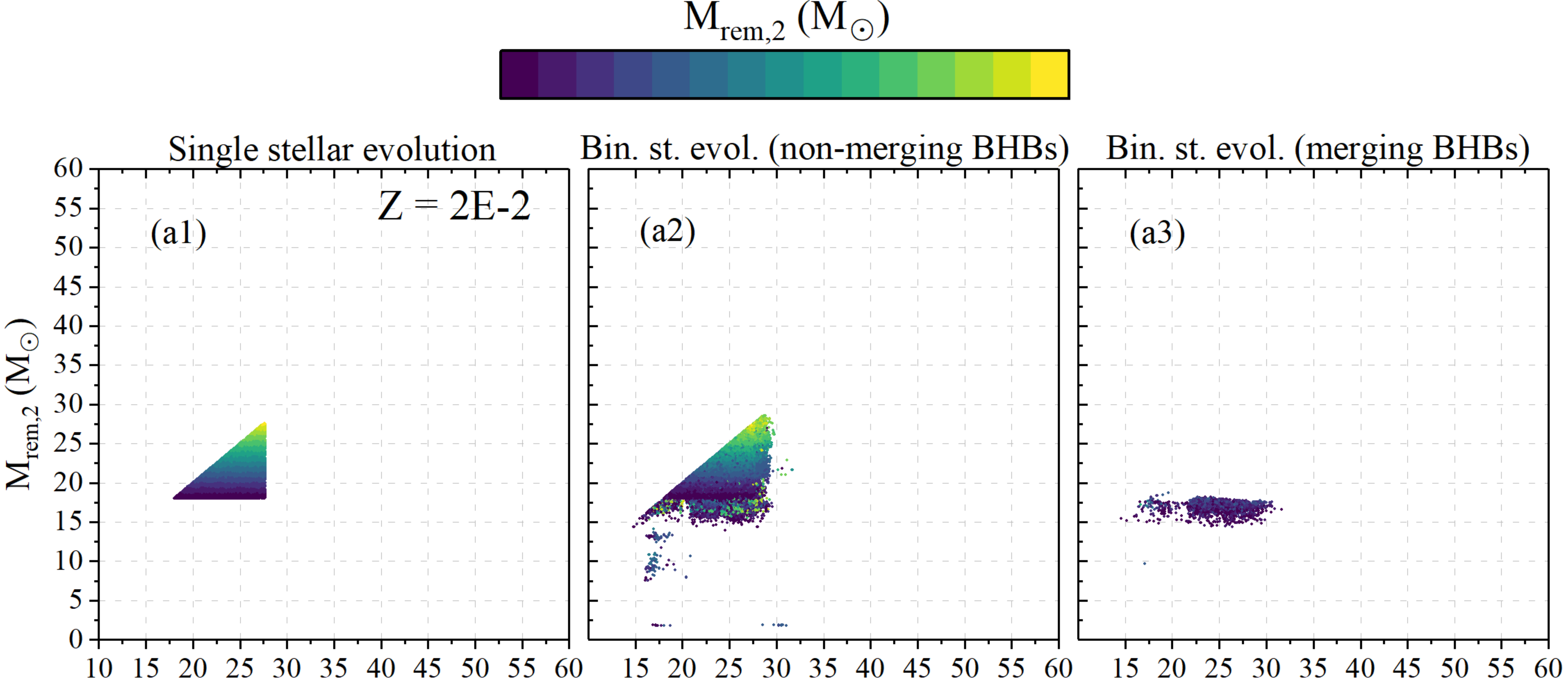}
	\includegraphics[width=0.95\hsize]{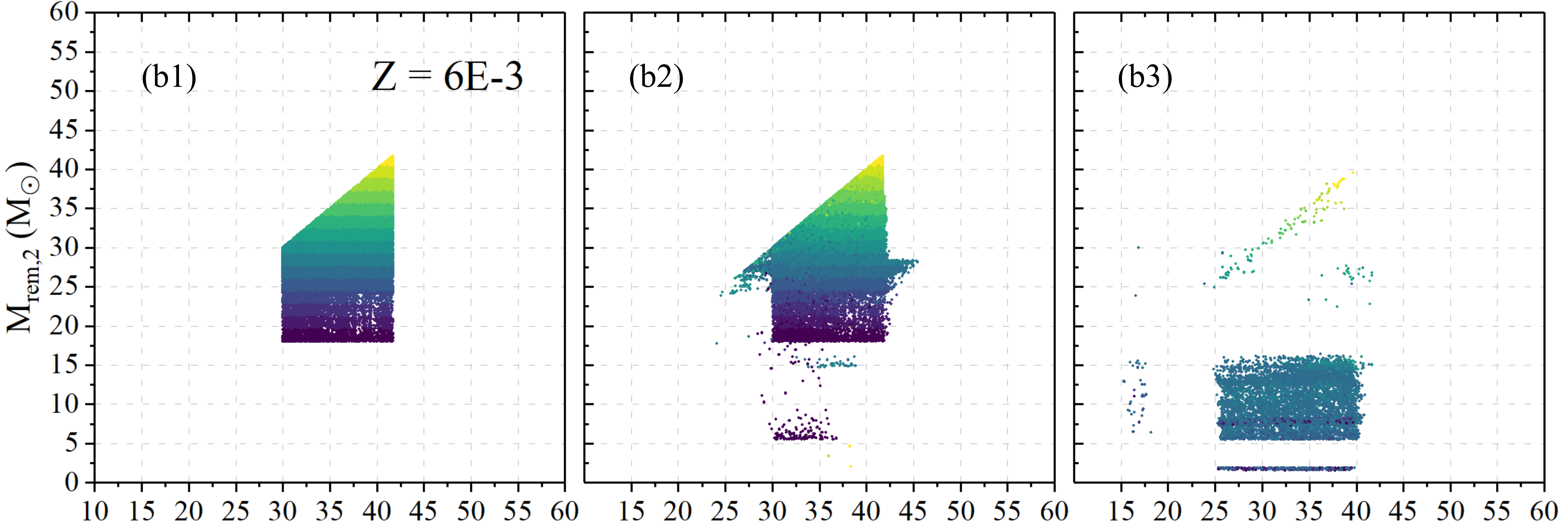}
	\includegraphics[width=0.95\hsize]{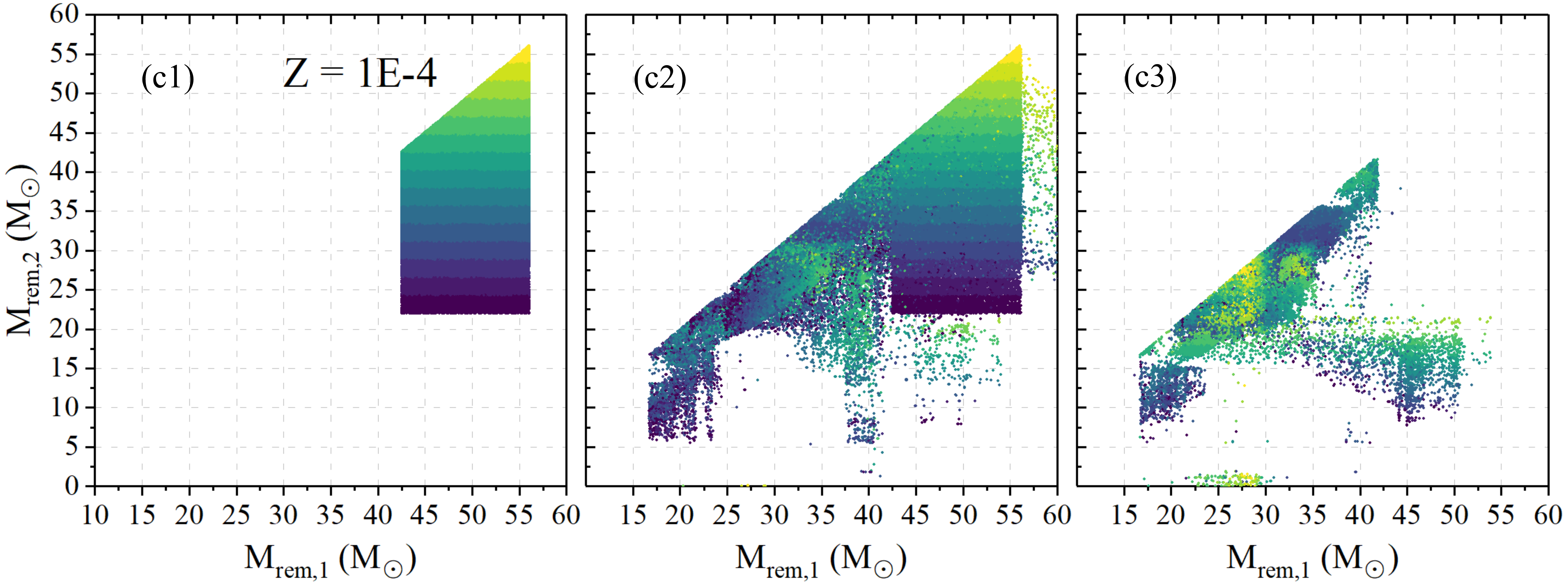}
	\caption{Mass of the less massive remnant ($M_{\rm rem,\,{}2}$) as a function of the mass of the more massive remnant ($M_{\rm rem,\,{}1}$) for the subset of double compact-object binaries that would form the heaviest BHs if we accounted only for single stellar evolution calculations. Left-hand column (labelled as \textit{1}): compact-object binaries that form if we account only for single stellar evolution processes; central column (labelled as \textit{2}): non-merging compact-object binaries that form if we account for both single and binary stellar evolution processes in our population-synthesis simulations; right-hand column (labelled as \textit{3}): same as in the central column but for merging binaries. Panels in the top row (labelled as \textit{a}): $Z=2\times 10^{-2}$; central row (\textit{b}): $Z=6\times 10^{-3}$; bottom row (\textit{c}): $Z=10^{-4}$. Colours show the value of the mass of the less massive remnant obtained from single stellar evolution (i.e. the values of $M_{\rm rem,\,{} 2}$ in the left-hand column).}
	\label{fig:m1m2_sse}
\end{figure*}

To better understand the evolution of massive BHBs, we extract from our simulations all the binaries that would have formed the most massive BHs if we had accounted only for single stellar evolution. We select them by looking at the zones of avoidance of massive merging BHs shown in the bottom row panels of Fig. \ref{fig:cobm1m2} and already described in Sec. \ref{sec:results}. Figure~\ref{fig:m1m2_sse} shows the mass of the less massive remnant as a function of the mass of the more massive remnant for such binary systems. The left-hand column shows the BHBs we obtain if we account only for single stellar evolution. The other two columns show the BHBs formed when including also binary stellar evolution processes. In particular, the central column shows the non-merging BHBs while the right-hand column shows the merging BHBs.

From Fig. \ref{fig:m1m2_sse} it is apparent that merging BHs tend to be lighter than we would have expected evolving their progenitors through single stellar evolution. The area filled by BHs in the left-hand column of Fig. \ref{fig:m1m2_sse} is mostly empty in the right-hand column.

Two massive progenitor stars may
\begin{itemize}
	\item merge during the MS phase, if they are born too close to each other ($a \lesssim 50\rsun{}$);
	\item evolve through no (or minor) mass-transfer episodes if they are born too far away from each other ($a \gtrsim$ few $10^3\rsun{}$);
	\item interact significantly with each other if $a\in \left[50\rsun{};  10^3\rsun{}\right]$.
\end{itemize}
In the first case, the stars merge and form one single massive star. In the second case, the progenitor stars do form a double compact-object but the remnants do not merge within a Hubble time because the semi-major axis is too large  (most of the systems in the central column of Fig. \ref{fig:m1m2_sse} belong to this category). In the third case, the progenitor stars may form a merging BHB (right-hand column of Fig. \ref{fig:m1m2_sse}).

At low metallicity (bottom row of Fig. \ref{fig:m1m2_sse}) stellar winds are quenched, therefore the heaviest BHs should come from progenitor stars with large radii ($\gtrsim 10^3 \rsun{}$) and massive Hydrogen envelopes ($\gtrsim 20\msun{}$). When such stars interact with each other, a stable Roche-lobe mass transfer phase and/or a CE evolution may significantly shrink the binary system so that $a \lesssim 10^2 \rsun{}$ and the BHs merge within a Hubble time. Still, most of the massive Hydrogen envelopes are lost during Roche-lobe overflow and CE, therefore the resulting BHs are significantly lighter than those formed considering only single stellar evolution. This effect is particularly strong at low $Z$ and for massive progenitor stars, that is for stars with massive Hydrogen envelopes.

It is also worth noting that a merging BHB with high mass ratio can form if a binary system evolves through a stable Roche-lobe mass transfer (or CE phase) when one of the two stars has already turned into a BH (panel \textit{c3} of Fig. \ref{fig:m1m2_sse} for $M_{\mathrm{rem,1}}\gtrsim 40\msun{}$ and $M_{\mathrm{rem,2}}\lesssim 20\msun{}$). In contrast, equal-mass merging BHs may form if the progenitor stars undergo a stable Roche-lobe mass transfer followed by a CE phase (less likely a double CE evolution)  and transform into two bare-He cores. Our stellar evolution prescriptions at $Z=10^{-4}$ predict that the maximum BH mass that  can result from a bare-He star is $\sim 40\msun{}$. This explains the cut-off at $M_{\mathrm{rem,1}}\simeq 40\msun{}$ observed in panel \textit{c3} of Fig.\ref{fig:m1m2_sse} and in panel \textit{b3} of Fig. \ref{fig:cobm1m2}.

At high metallicity the situation is quite different. From single stellar evolution calculations we know that the heaviest BHs should form from WR stars, that is stars with small radii (few $\rsun{}$) and without a Hydrogen envelope (lost through stellar winds). In our models, at $Z=2\times 10^{-2}$, most of massive progenitors have always $R \lesssim 10^2\rsun{}$ during their life (cf. panel \textit{b1} of Fig. \ref{fig:comparison}), therefore they are unlikely to interact with each other because they are quite small. This implies that a BHB formed from such progenitors unlikely becomes tight enough to merge within a Hubble time. To obtain tighter BHBs we need lighter progenitor stars, that is stars that expand significantly before turning into bare-He stars, so that they can evolve through a CE phase. This also explains why in panel \textit{a3} of Fig. \ref{fig:m1m2_sse} we have only dark-blue points, that is quite light merging BHBs ($M_{\mathrm{rem,2}}\lesssim 20\msun{}$) that come only from relatively light progenitors ($M_\mathrm{ZAMS}\lesssim 75\msun{}$).

At $Z=6\times 10^{-3}$ the situation is intermediate. Most of the heaviest BHs are still expected to come from WR stars but their progenitors may reach quite large radii ($\gtrsim 10^2\rsun{}$) before turning into bare-He cores. Some of these progenitors may still evolve through a CE phase but the binary system cannot shrink significantly because the Hydrogen envelopes are too light ($ <10\msun{}$). This also explains why in panel \textit{b3} of Fig. \ref{fig:m1m2_sse} we have only very few points with $M_{\mathrm{rem,2}}\gtrsim 25\msun{}$. Merging BHBs can still form after a CE evolution provided that the shared envelope is quite massive ($\gtrsim 10\msun{}$), but in this case at least one of the two BHs must be quite light, as already discussed for the low-$Z$ case.

\subsection{Number of BH mergers}
Stellar winds and stellar radii are crucial ingredients to understand how the number of merging BHs depends on metallicity. At high $Z$,  the semi-major axis of binary stars may easily increase because of strong stellar winds, therefore BHBs tend to have larger separations. Furthermore, the most massive stars ($M_{\mathrm{ZAMS}}\gtrsim 75 \msun{}$ at $Z=2\times 10^{-2}$) lose all their Hydrogen envelope via stellar winds without turning into supergiants. This means that metal-rich stars have also less chances to interact with each other because WR stars have quite small radii. Even though lighter stars ($M_{\mathrm{ZAMS}}\lesssim 75 \msun{}\msun{}$ at $Z=2\times 10^{-2}$) may undergo a CE phase, the shared envelope is likely quite light (because stellar winds removed a large fraction of the envelope), therefore metal-rich stars  have also less mass reservoir that can be used to shrink binary systems.

\begin{figure}
	
	\includegraphics[width=\hsize]{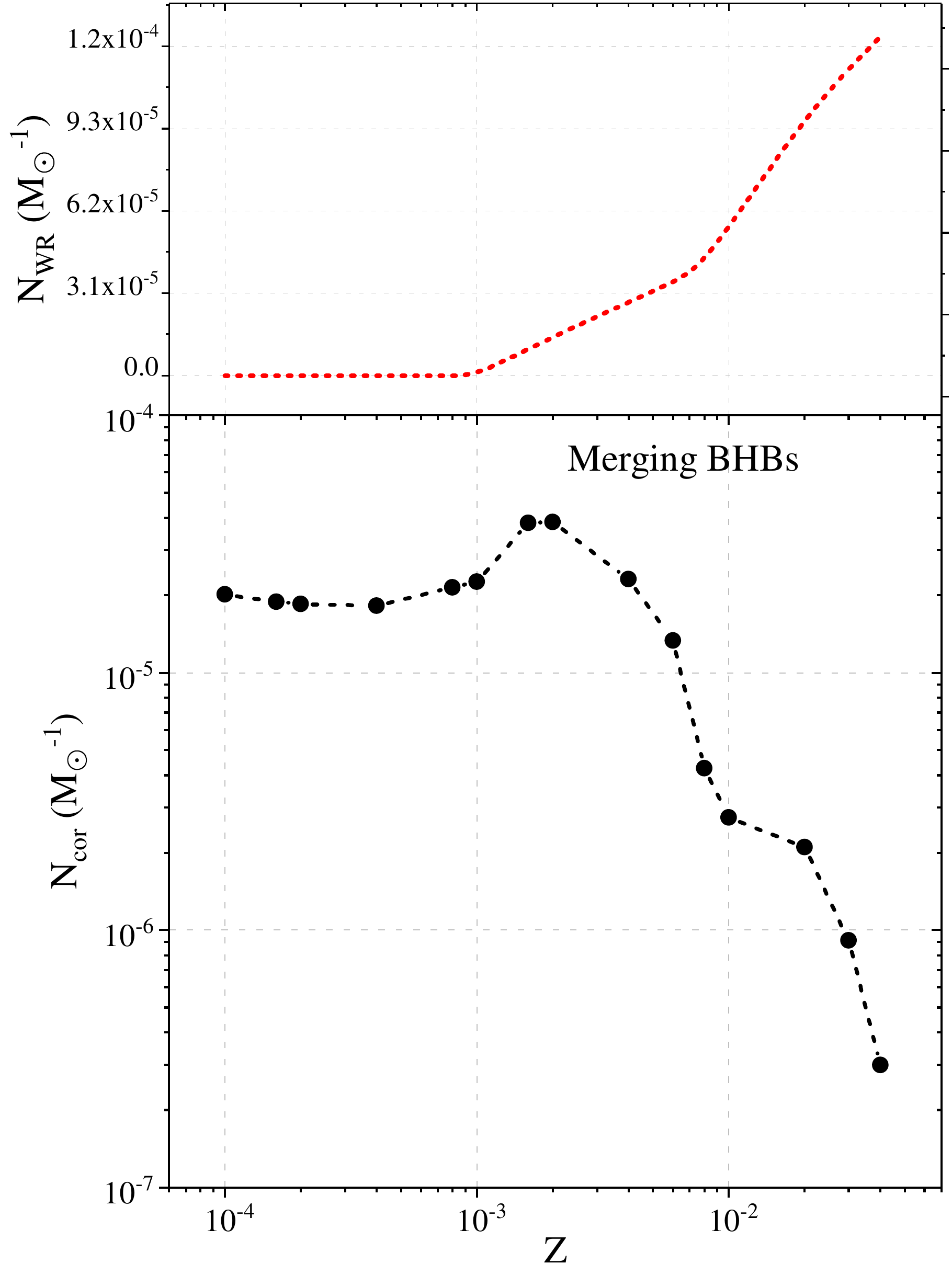}
	\caption{Top panel: number of WR stars per unit mass predicted by the \sevn{} code, as a function of metallicity. The data have been obtained for the stellar population used in our simulations and considering only single stellar evolution processes. Bottom panel: number of merging BHBs per unit mass as a function of metallicity.}
	\label{fig:N_vs_Z}
\end{figure}

For these reasons, we expect a higher number of BH mergers at low metallicity, where stellar winds are quenched and stars can reach larger radii and retain more massive envelopes. Figure \ref{fig:N_vs_Z} confirms the expectations. It shows the number of merging BHBs per unit stellar mass in our simulations ($N_{\mathrm{cor,BHB}}$, bottom panel) and the number of WR stars per unit stellar mass (top panel) predicted by our single stellar evolution models, as a function of metallicity.

We compute $N_{\mathrm{cor,BHB}}$ following the formula given in \citet{giacobbo2018}:
\begin{equation}
\label{eq:ncorbbh}
N_{\mathrm{cor,BHB}} = f_{\mathrm{bin}}f_{\mathrm{IMF}}\frac{N_{\mathrm{mergers,BHB}}}{M_{\mathrm{tot}}}
\end{equation}
where $N_{\mathrm{mergers,BHB}}$ is the number of merging BHBs, $M_{\mathrm{tot}}$ is the total initial mass of the simulated stellar population, $f_{\mathrm{IMF}}$ corrects for the fact that we have simulated only stars with ZAMS mass $M_{\rm ZAMS}\ge{}10\msun{}$ ($f_{\mathrm{IMF}} = 0.137$), and $f_{\mathrm{bin}}$ is a correction factor which accounts for the fact that all stars in our sample are members of binary systems. To compute $N_{\mathrm{cor,BHB}}$ we assume that only 50\% of stars are binaries \citep{sana2013},  that is $f_{\mathrm{bin}}=0.5$. Figure \ref{fig:N_vs_Z} shows a peak of BH mergers at $Z\simeq 3\times 10^{-3}$, which corresponds to the lowest metallicity at which massive single stars can evolve into WR stars (top panel, dashed red curve). 

Figure~\ref{fig:N_vs_Z} also shows a mild decrease of the number of merging BHBs at $Z\lesssim 2\times 10^{-3}$. At $Z=2\times 10^{-3}$ we have a factor of $\sim 3$ more merging BHBs than at $Z=10^{-4}$. The onset of PISNe plays only a minor role: PISNe disrupt the progenitors of heavy BHs before they can form a remnant, reducing the number of BHs at low metallicity; on the other hand, only the most massive stars ($M_{\rm He}\gtrsim{}60$ M$_\odot$) explode as PISNe, thus their impact on $N_{\mathrm{cor,BHB}}$ is negligible.

The decrease of $N_{\mathrm{cor,BHB}}$ at $Z\le{}2\times{}10^{-3}$ mainly happens because  we form significantly more BHBs at $Z=2\times 10^{-3}$ than at $Z=10^{-4}$. In our simulations, binary stars with members with $M_{\mathrm{ZAMS}}\in\left[15;30\right]\msun{}$  (which produce a large fraction of all double compact objects) form more or less the same number of double compact objects at $Z=2\times 10^{-3}$ and $Z=10^{-4}$ ($\sim 1.5\times{}10^5$ double compact objects). The vast majority of these double compact objects at $Z=10^{-4}$ are BH-NS binaries ($\sim{}1.1\times{}10^5$), while the number of BHBs is $2\times{}10^4$ and only $\sim{}1000$ of them merge within a Hubble time. We obtain the same result if we evolve the same systems considering only single stellar evolution calculations.

In contrast, at  $Z=2\times 10^{-3}$, we find $\sim 1.1\times{}10^5$ BHBs ($\sim 2\times{}10^4$ of them merge within a Hubble time), and only $\sim 4\times{}10^4$ BH-NS systems.

The evolution of the considered progenitor stars is similar at both $Z=2\times 10^{-3}$ and $Z=10^{-4}$: double compact objects form after a CE phase involving a BH (formed from the primary star) and the secondary star. The difference is that at $Z=2\times 10^{-3}$, the primary star, before turning into a BH, fills the Roche lobe and the system evolves through a stable mass transfer phase. In this case, the mass transferred from the primary star is enough to let the secondary star form a BH instead of a NS, after the CE evolution. In contrast, at $Z=10^{-4}$, progenitor stars have smaller radii, therefore the considered binary systems do not evolve though a stable mass-transfer phase before entering CE. 

It is also worth noting that at $Z=2\times 10^{-3}$ the secondary star undergoes the CE evolution when it is in the core-Helium burning phase, whereas at $Z=10^{-4}$ the star has already formed a CO core and it has Helium and Hydrogen in the outer shells.

The reason of this difference is that, according to {\sc PARSEC} evolutionary tracks\footnote{Stellar radii are crucial to understand when the secondary star evolves through a CE phase. In this respect it is important to remind that the evolution of massive stars in the HR diagram, after central Hydrogen burning, strongly depends on the details of the input physics \citep{Chiosi_Summa70,tang2014}. Stars with relatively lower mass undergo a very similar evolution to intermediate mass stars, reaching central He ignition in the red supergiant region. In more massive stars, however, it is possible that central Helium ignition happens already in the blue/yellow supergiant phase.}, stars with $M_{\mathrm{ZAMS}}\in\left[15;30\right]\msun{}$ and metallicity  $Z=2\times 10^{-3}$ ignite Helium as red supergiant stars. Such stars have quite large radii and likely undergo a CE phase during the He-core burning phase.

In contrast, stars with the same mass ($M_{\mathrm{ZAMS}}\in\left[15;30\right]\msun{}$) and metallicity $Z=10^{-4}$ ignite Helium as yellow/blue supergiants, which means that they are not large enough to evolve through a CE phase at that stage. Such stars can enter CE only when they turn into red supergiant stars, that is, when they have already formed a CO core.

\subsection{Local merger rate density}
We use the results of our simulations to estimate the local merger rate density of BHBs ($R_{\mathrm{loc,BHB}}$) and we compare it with the rate inferred from the LIGO-Virgo data.

To calculate the merger rate density of BHBs in the local Universe ($R_{\rm BHB}$) we adopt the simple analytic calculation described in Section 3.5 of \cite{giacobbo2018b}:
\begin{equation}\label{eq:rate}
R_{\rm BHB} = \frac{1}{H_0\,{}t_{\rm lb}(z=0.1)} \int_{z_{\rm max}}^{z_{\rm min}} \frac{f_{\rm loc}(z)~ {\rm SFR}(z)}{(1+z)\,{}\left[\Omega_{\rm M}\,{}(1+z)^3+\Omega_\lambda{}\right]^{1/2}}\,{}dz,
\end{equation}
where ${\rm SFR}(z)$ is the star formation rate density as a function of redshift (we adopt the fitting formula provided in \citealt{madau2014}), $t_{\rm lb}(z=0.1)$ is the look back time at redshift $z=0.1$, $f_{\rm loc}(z)$ is the fraction of binaries which form at redshift $z$ and merge in the local Universe (defined as $z\le{}0.1$), $z_{\rm max}=15$, $z_{\rm min}=0$, while $H_0$, $\Omega_{\rm M}$ and $\Omega_\lambda{}$ are the cosmological parameters (for which we adopt values from \citealt{planck2016})\footnote{Note that equation~\ref{eq:rate} is the same as equation~10 of \cite{giacobbo2018b}, but is written as an integral rather than a summation.}.


We calculate $f_{\rm loc}(z)$ from $N_{\mathrm{cor,BHB}}$ (equation~\ref{eq:ncorbbh}), assuming that all stars in the same redshift bin have the same metallicity. We compute the metallicity at a given redshift as $\log Z\left(z\right)/\zsun{} = -0.19z$ if $z\leq1.5$ and $\log Z\left(z\right)/\zsun{} = -0.22z$ if $z>1.5$. This formula comes from abundance measurements of a large sample of high-redshift damped Ly$\alpha$ systems \citep{rafelski2012}, but re-scaled to have $Z\left(z=0\right)=\zsun{}$, consistent with the Sloan Digital Sky Survey data \citep{gallazzi2008}.

Our model predicts $R_{\mathrm{loc,BHB}}\simeq 90\,\mathrm{Gpc}^{-3}\,\mathrm{yr}^{-1}$, consistent with 
the BHB merger rate inferred from LIGO-Virgo data {   (24 -- 112 $\mathrm{Gpc}^{-3}\,\mathrm{yr}^{-1}$ \citealt{catalog2018,poprate2018}).}

\subsection{Formation of massive single BHs}
We have shown in Section \ref{sec:results} (Fig. \ref{fig:nbh}) that while the mass distribution of BHs that are members of double compact objects is similar to the one obtained from single stellar evolution, the mass distribution of single BHs is quite peculiar, especially at low metallicity. We know that PPISNe significantly enhance mass loss from massive progenitor stars and PISNe  disrupt  massive stars before they can form a heavy BH. Thus, from single stellar evolution, we do not expect to form BHs with mass $\gtrsim 60\msun{}$ \citep{spera2017}.

In contrast, if we account for binary evolution processes, we can form single BHs with mass up to $\sim 65$, $ 90$, and $145 \msun{}$ at $Z=2\times 10^{-2}$, $6\times10^{-3}$, and $10^{-4}$, respectively. In our simulations, such heavy BHs may form from the merger of two MS stars when one of them is at the end of the MS phase. In this case, the \sevn{} code assumes that the merger product is also at the end of the MS phase. Even if the merger product has a significant amount of Hydrogen, we assume that most of it is part of the envelope, therefore it will not be transformed into Helium by nuclear reactions. This implies that the mass of the Helium core of the merger product may be lower than the limit for a PISN to occur. Thus, the merger product can form a BH by direct collapse, and such BH can be very massive, considering the large mass of the Hydrogen envelope.


This effect is more pronounced at low metallicity where stellar winds are not strong enough to remove the massive Hydrogen envelope of the merger product. Since such massive BHs are single and very rare ($\lesssim 0.1\%$ of the total number of BHs at $Z=10^{-4}$) they do not play a major role in binary population-synthesis simulations. In contrast, they can be very important if they form in star clusters, where they have a high chance to acquire a companion through dynamical exchanges, so that they possibly become loud GW sources \citep{portegieszwart2000,mapelli2016,askar2017}.

{   
\subsection{Comparison with \combine{}}
Recently, \citet{kruckow2018} used a new grid-based population synthesis code (\combine{}) to study the formation and evolution of double compact-object binaries. In this section we discuss the main differences between \sevn{} and \combine{}. 

Both codes interpolate look-up tables to evolve the physical parameters of single stars. Furthermore, they both use similar criteria to jump on new tracks whenever a star has accreted (donated) a significant amount of mass from (to) its companion. The main differences in the interpolation scheme are:
\begin{enumerate}
	\item \combine{} interpolates only on the mass variable while \sevn{} interpolates also over different metallicities. This means that with \sevn{} we can evolve stars at any metallicity between $Z=10^{-4}$ and $Z=4\times 10^{-2}$;
	
	\item \combine{} uses linear weights to interpolate tracks (see eqs. A1 and A2 of \citealt{kruckow2018}), while in \sevn{} we use more sophisticated weights that significantly improve interpolation errors (see eqs. A3 and A4 and \citet{spera2017} for details);
	
	\item to calculate the interpolation time for the stellar tracks, \combine{} uses the ratio between the current age of a star and its total lifetime (see eq. A6 of \citealt{kruckow2018}), while in \sevn{} we improve the accuracy af the interpolation by using the relative age of a star with respect to its current evolutionary phase (see Sec. 2.1.1 and Appendix A).
	
\end{enumerate}

The main difference in terms of scientific results is that \combine{} matches the low-mass GW events (GW151226 and GW170608) only at high metallicity (see Fig. 15 of \citealt{kruckow2018}), while with \sevn{} we can form such events at all metallicities (see Fig. \ref{fig:cobm1m2}). 
This happens because the two codes use different prescriptions for the formation of BHs. 
In \combine{}, a BH forms if the final  CO core mass of a star is $>6.5~\msun{}$ and the BH mass is calculated assuming a fixed amount of fallback ($80\%$ of the mass of the He envelope). This implies that, at high metallicity, \combine{} forms BHs with mass $\gtrsim 6.5\msun{}$ (see Fig. 2 of \citet{kruckow2018}). At low metallicity, stellar winds are quenched, therefore stars have more massive He envelopes and the minimum BH mass is larger ($\gtrsim 10\msun{}$). In contrast, in our paper we adopt the rapid SN explosion model \citep{fryer2012}, which predicts variable fallback and does not distinguish a priori between NSs and BHs. We impose that all compact objects with mass $\geq 3\msun{}$ are BHs while the others are NSs. As a consequence, in \sevn{} we can form smaller BHs (down to $\sim 5\msun{}$) from the collapse of stars with final  CO core masses of $\sim 4\msun{}$, and with small fallback fractions ($\sim 5\%$). Furthermore, our minimum BH mass is quite insensitive to metallicity because BHs of $\sim 5\msun{}$ form from stars with $M_{\rm ZAMS}\simeq 25\msun{}$ whose final physical parameters mildly depend on metallicity.

The difference in the maximum BH mass at high metallicity ($\sim{}11$ and $35$ M$_\odot$ in ComBinE and SEVN, respectively) is also a consequence of the different SN explosion prescriptions adopted by the two codes. At low metallicity, both \sevn{} and \combine{} form BHs with mass up to $\sim 60\msun{}$ (see Fig. 7 of \citealt{kruckow2018} and our Fig. \ref{fig:cobm1m2}). Still, while the $60\msun{}$ cut-off in \sevn{} comes from PPISNe and PISNe, the latter are not included in \combine{}. The inclusion of PPISNe and PISNe would significantly reduce the maximum BH mass obtained by \combine{} at low metallicity.

Other differences come from  binary stellar evolution prescriptions. For example, the prescriptions for BH natal kicks are quite different in the two codes.  \combine{} adopts a flat distribution of BH natal kicks between $0$ and $200\,\mathrm{km/s}$, while \sevn{} assumes the \citet{hobbs2005} distribution ($\sigma_{\rm 1D} = 265\,\mathrm{km/s}$) scaled by the amount of fallback mass (see Sec. 2.2). Furthermore, both codes adopt the $\alpha{}$$\lambda$-formalism for the common envelope phase, but \combine{} calculates the $\lambda$ parameter directly from the stellar structure, while \sevn{} still assumes a fixed $\lambda$ (we use $\lambda=0.1$ for this paper). We will improve this aspect in the next version of \sevn{}. Finally, wind accretion and tidal evolution are not included in \combine{}. Simultaneous circularization is assumed by \combine{} when the binary is at the onset of Roche lobe mass transfer. Furthermore, in \combine{}, the donor star is always assumed to transfer its envelope entirely to the accretor, while in \sevn{} we limit mass loss by taking into account of several factors (see our eqs. 7-8). 

Overall, \sevn{} and \combine{} both share the same novel approach (which consists in interpolating stellar evolution from look-up tables), but they have also important differences in the interpolation algorithm, in the implementation of binary evolution processes and, more importantly, in the formation of compact remnants. This leads to a significantly different mass spectrum of merging BHs, although the results of both codes are still fairly consistent with GW detections.

}


\section{Conclusions}
We investigated the statistics of BHBs using a new version of the \sevn{} population-synthesis code (\citealt{spera2015,spera2017}). To compute the evolution of physical stellar parameters, \sevn{} interpolates a set of tabulated stellar evolutionary tracks on-the-fly. The default look-up tables come from the \parsec{} stellar evolution code (\citealt{bressan2012,chen2015}). \sevn{} also includes five different models for core-collapse SNe and prescriptions to model PPISNe and PISNe.  We updated the \sevn{} code by adding binary stellar evolution processes (wind mass transfer, Roche-lobe mass transfer, common envelope, mergers, tides, and GW decay). We also developed a novel algorithm to couple the interpolation of the look-up tables with the binary stellar evolution formulas (supplementary material, Appendix A). 

We used the new version of the \sevn{} code to run 15 sets of simulations with 15 different metallicities ($Z\in \left[10^{-4};4\times 10^{-2}\right]$). Each simulation evolves a sample of $10^7$ binary systems until all stars have turned into compact remnants.

We found that the mass distribution of BHs which are members of compact-object binaries is quite similar to the one obtained considering only single stellar evolution calculations (Figs. \ref{fig:nbh} and \ref{fig:spectrum002}). The maximum BH mass in binary systems is $\sim 30$, $45$ and $55\msun{}$ at metallicity $Z=2\times 10^{-2}$, $6\times 10^{-3}$, and $10^{-4}$, respectively.

In contrast, the mass distribution of single BHs is very different. We form single BHs with mass up to $\sim 65$, $90$, and $145\msun{}$ at metallicity $Z=2\times 10^{-2}$, $6\times 10^{-3}$, and $10^{-4}$, respectively. Such massive BHs fall right into the BH mass-gap ($60 - 120\msun{}$) produced by PPISNe and PISNe (\citealt{belczynski2016pair,spera2017,woosley2017}). These heavy BHs come from  the merger of two MS stars when one of the two stars is at the end of MS. While these BHs are very rare ($\lesssim{}0.1$ \% of all BHs at $Z=10^{-4}$), they may be important if they form in star clusters, where they have a high chance to acquire a companion via dynamical exchanges and to become GW sources.

In our simulations, the BHBs hosting the heaviest BHs are unlikely to merge within a Hubble time, in agreement with \cite{giacobbo2018}. We found no merging BHBs with both BHs more massive than $\sim 18$, $25$, and $40\msun{}$ at $Z=2\times 10^{-2}$, $6\times 10^{-3}$ and $10^{-4}$, respectively. Stellar radii and Hydrogen envelopes play a crucial role to explain why the most massive BHs do not merge (e.g. panel \textit{b1} of Fig. \ref{fig:comparison}). In particular, at low metallicity the progenitors of the heaviest BHs reach quite large radii ($\gtrsim 10^3 \rsun{}$) and retain massive Hydrogen envelopes ($\gtrsim 20\msun{}$). When such massive stars evolve through a stable Roche-lobe mass transfer or a CE evolution, the orbit shrinks and the massive Hydrogen envelopes are lost. Thus, the BHs formed from massive metal-poor progenitors likely merge within a Hubble time via GWs but they are also quite light, because of the mass lost during CE.

At high metallicity, the progenitors of the heaviest BHs are WR stars, that is stars with small radii (a few $\rsun{}$) and no Hydrogen envelopes. This means that such progenitors are unlikely to enter CE and to get close enough to each other to merge within a Hubble time (Figs. \ref{fig:cobm1m2} and \ref{fig:m1m2_sse}). 

For similar reasons, merging BHBs form more efficiently from metal-poor than from metal-rich progenitors: we expect BH mergers to be two orders of magnitude more frequent from stars with $Z\leq{}2\times{}10^{-3}$ than from solar metallicity stars. This happens because metal-rich stars tend to have small radii and to develop light envelopes (if any), because of the strong stellar winds. With such small radii and envelopes, they can hardly enter a CE phase and they fail to reduce their orbital periods.

The number of merging BHBs is maximum for metallicity $Z\sim{}2\times{}10^{-3}$, while it drops at higher metallicity and it decreases by a factor of $\sim{}3$ at lower metallicity. If the star metallicity is very low ($Z<10^{-3}$), we tend to form more BH-NS binaries than BHBs, therefore the number of merging BHBs decreases slightly with respect to $Z\sim{}2\times{}10^{-3}$. This happens because, at $Z\sim{}2\times{}10^{-3}$, primary stars with $M_{\mathrm{ZAMS}}\in\left[20;30\right]\msun{}$ can evolve though a stable mass-transfer phase and the secondary star ($M_{\mathrm{ZAMS}}\in\left[15;25\right]\msun{}$) may acquire enough mass to form a BH instead of a NS. In contrast, at very low metallicity the mass-transfer phase is more unlikely to happen because stars have smaller radii.

Finally, we compared our results against LIGO-Virgo detections. In our simulations, we found that merging BHBs with masses consistent with the low-mass GW events {   (GW151226, GW170608, and GW151012) can form at all metallicities. In contrast, merging BHs consistent with the GW events with primary BH mass $>30$ M$_\odot$ form only from metal-poor progenitors (Figs. \ref{fig:cobm1m2}, \ref{fig:chirp}, \ref{fig:ratio}). We also found that it is unlikely to form merging BHBs with masses consistent with GW170729 (i.e the GW event with the heaviest BHs).} We do not form merging BHs with $M_{\mathrm{chirp}}\gtrsim 37\msun{}$, independently of metallicity. {   This is in good agreement with O1 and O2 detections, which suggest a dearth of merging BHs with mass $M_{\rm BH}>45$ M$_\odot$ \citep{poprate2018}.}

The merger rate of BHBs in the local Universe estimated from our models is $\sim90\,\mathrm{Gpc}^{-3}\mathrm{yr}^{-1}$, consistent with the BHB merger rate inferred from LIGO-Virgo data {   (24 -- 112 $\mathrm{Gpc}^{-3}\,\mathrm{yr}^{-1}$, \citealt{catalog2018,poprate2018}).}

Our results confirm that stellar winds, stellar radii and binary evolution processes (especially mass transfer and common envelope) are key ingredients to understand the statistics of merging BHs across cosmic time. Our new version of the \sevn{} code is uniquely suited to investigate this topic. {   In a follow-up paper we will consider stellar metallicity down to $Z\sim{}10^{-7}$ and ZAMS masses up to $\sim{}350$ M$_\odot$. The next step is to use \sevn{} in combination with $N$-body simulations to study the role of stellar dynamics on the formation and evolution of BHBs.}

\section*{Acknowledgments}
We acknowledge the ``Accordo Quadro INAF-CINECA (2017)'' and the CINECA-INFN agreement for the availability of high performance computing resources and support. MS acknowledges funding from the European Union's Horizon 2020 research and innovation programme under the Marie-Sklodowska-Curie grant agreement No. 794393. MM acknowledges financial support  by the European Research Council for the ERC Consolidator grant DEMOBLACK, under contract no. 770017. AAT acknowledges support from JSPS KAKENHI Grant Number 17F17764. NG acknowledges financial support from Fondazione Ing. Aldo Gini and thanks the Institute for Astrophysics and Particle Physics of the University of Innsbruck for hosting him during the preparation of this paper. This work benefited from support by the International Space Science Institute (ISSI), Bern, Switzerland,  through its International Team programme ref. no. 393 {\it The Evolution of Rich Stellar Populations \& BH Binaries} (2017-18). MS thanks the Aspen Center for Physics, which is supported by National Science Foundation grant PHY-1607611, where part of this work was performed.

\appendix

\section{Interpolation}\label{app:interpolation}
\subsection{Single stars}\label{subapp:singlestars}
To evolve a star $s$ with ZAMS mass $M_{\mathrm{ZAMS,s}}$ and metallicity $Z_s$ at time $t$, in \sevn{} we use four interpolation tracks from the look-up tables. Two of them have metallicity $Z_1 = Z_s - \Delta Z$ and ZAMS masses $M_{\mathrm{ZAMS,1}}=M_{\mathrm{ZAMS,s}}-\Delta M$ and $M_{\mathrm{ZAMS,2}}=M_{\mathrm{ZAMS,s}}+\Delta M$, respectively, where $\Delta M$ is the step of the mass grid of the look-up tables at metallicity $Z_1$ and $\Delta Z$ is the step of the metallicity of the look-up tables. The other two interpolation tracks have ZAMS masses $M_{\mathrm{ZAMS,1}}$ and $M_{\mathrm{ZAMS,2}}$ and metallicity $Z_2 = Z_s + \Delta Z$. To begin with, we calculate the percentage of life $\Theta_{\rm p}$ of the star $s$ on its macro-phase $p$ (see equation 1). Then, we use the interpolation tracks to calculate
\begin{equation}
\label{eq:masscalc}
M_{s,Z_j}\left(t\right) = \beta_1 M_{1,Z_j}\left(t_{1,Z_j}\right) + \beta_2 M_{2,Z_j}\left(t_{2,Z_j}\right)
\end{equation}where 
\begin{equation}
t_{i,Z_j}\equiv \Theta_{\rm p}\left(t_{\mathrm{f,p,i,Z_j}} - t_{\mathrm{0,p,i,Z_j}}\right) + t_{\mathrm{0,p,i,Z_j}},
\end{equation}
\begin{equation}
\beta_1\equiv\frac{M_{\mathrm{ZAMS,1}}\left(M_{\mathrm{ZAMS,2}}-M_{\mathrm{ZAMS,s}}\right)}{M_{\mathrm{ZAMS,s}}\left(M_{\mathrm{ZAMS,2}}-M_{\mathrm{ZAMS,1}}\right)},
\end{equation}
\begin{equation}
\beta_2\equiv\frac{M_{\mathrm{ZAMS,2}}\left(M_{\mathrm{ZAMS,s}}-M_{\mathrm{ZAMS,1}}\right)}{M_{\mathrm{ZAMS,s}}\left(M_{\mathrm{ZAMS,2}}-M_{\mathrm{ZAMS,1}}\right)},
\end{equation}
$i,j\in\left[1,2\right]$ and $t_{\mathrm{0,p,i,Z_j}}$ $\left(t_{\mathrm{f,p,i,Z_j}}\right)$ is the starting (end) time of the macro-phase $p$ of the star with ZAMS mass $M_i$ at metallicity $Z_j$.

The use of the $\beta_1$ and $\beta_2$ weights allows us to keep the interpolation error below 1\% with respect to the original PARSEC evolutionary tracks and to include less points in the look-up tables. We refer to \citet{spera2017} for more details on the weights.
We calculate the interpolated value of the mass of the star $s$ at time $t$ as
\begin{equation}
M_s\left(t\right) = \gamma_1 \,{}M_{s,Z_1}\left(t\right) + \gamma_2\,{} M_{s,Z_2}\left(t\right),
\end{equation}
where $\gamma_1 \equiv \frac{Z_s-Z_1}{Z_2-Z_1}$ and $\gamma_2 \equiv \frac{Z_2-Z_s}{Z_2-Z_1}$.

\subsection{Binary stars}\label{subapp:binarystars}
Every time a star has accreted (donated) a significant amount of mass $\Delta m$ from (to) its companion, the code looks for new interpolation tracks in the look-up tables. We avoid to jump to new tracks when $\Delta m$ is small, thus we impose that a change of track can occur only if 
$\Delta m>\gamma_{m}\,{}M$,
where $M$ is the total mass of the star, $\gamma_{m}$ is a parameter with typical value of $\sim 0.01$, and $\Delta m$ is the mass exchange (loss or gain). 
The rules  governing a change of track depend primarily on the star's macro-phase. 

Hereafter, we call {\em old star} the star that is still on the old track while we use {\em new star} to refer to the star that has moved to a new track.

\subsubsection{Stars in the H phase}
To change the evolutionary track for stars in the H phase we require that the new star has the same percentage of life ($\Theta_{\rm H}$) and the same total mass of the old star. We define 

\begin{equation}
\Theta_{\rm H}\equiv\frac{t_{\mathrm{loc}}}{t_{\mathrm{HeS}}},
\label{eq:perclife}
\end{equation}

where $t_{\mathrm{loc}}$ is the time on the stellar track and $t_{\mathrm{HeS}}$ is the starting time of the He phase. We also check that 

\begin{equation}
\frac{\left|M_1-M_0\right|}{M_0}<\epsilon{}_1,
\label{eq:Hmassvar}
\end{equation}

where $M_1$ is the mass of the new star, $M_0$ is the mass of the old star, and $\epsilon{}_1$ is a parameter with a typical value of $\sim 10^{-3}$. To find the new track we implemented an iterative algorithm. We assume that the new track will not be far from the one with 

\begin{equation}
M_{\mathrm{ZAMS,new}} = M_{\mathrm{ZAMS,old}} \pm{} \Delta m.
\label{eq:tempzams}
\end{equation}
The sign in the above equation is $-$ ($+$) if the star is a donor (accretor).

Thus, we choose the interval $M_{\mathrm{ZAMS}}\in\left[M_{\mathrm{ZAMS,1}}, M_{\mathrm{ZAMS,2}}\right]$ as the fiducial range to find the new stellar track, where

\begin{equation}
M_{\mathrm{ZAMS,1}} = M_{\mathrm{ZAMS,old}} \pm{} \xi_{\mathrm{low}}\Delta m,
\label{eq:firsttrack}
\end{equation}

\begin{equation}
M_{\mathrm{ZAMS,2}} = M_{\mathrm{ZAMS,old}} \pm{} \xi_{\mathrm{high}}\Delta m.
\label{eq:secondtrack}
\end{equation}

The values of the parameters $\xi_{\mathrm{low}}$ and $\xi_{\mathrm{high}}$ are chosen by experiment and have typical values of $\sim 0.2$ and $\sim 1.2$, respectively. 
As starting values, the algorithm evaluates the mass of the star at times $t_1=p\,{}t_{\mathrm{HeS,1}}$ and $t_2=p\,{}t_{\mathrm{HeS,2}}$, where the subscripts 1 and 2 refer to the tracks with ZAMS masses $M_{\mathrm{ZAMS,1}}$ and $M_{\mathrm{ZAMS,2}}$, respectively. 
The next iterations are given by

\begin{eqnarray}
M_{0} - m\left(t_{n-1}\right) = \frac{m\left(t_{n}\right) - m\left(t_{n-1}\right)}{M_{\mathrm{ZAMS,}n} - M_{\mathrm{ZAMS,}n-1}}\nonumber{}\\
\times{}\left(M_{\mathrm{ZAMS,}n+1} - M_{\mathrm{ZAMS,}n-1}\right),
\label{eq:iteration}
\end{eqnarray}
where $n=0,1,2...,n_{\mathrm{max}}$, $M_{\mathrm{ZAMS,}n+1}$ is the ZAMS mass of the new candidate track and $m\left(t_{n}\right)$ is the mass of the star at time $t_{n}$ (here $n$ is the number of iterations). The algorithm stops if equation~(\ref{eq:Hmassvar}), with $M_1=m\left(t_{n+1}\right)$, is verified or, in any case, if $M_{\mathrm{ZAMS,}n+1}$ goes beyond the tracks included in the look-up tables, or if a maximum of $n_{\mathrm{max}}=$ \verb'MAX_iterations_H' iterations is reached. The default value of the \verb'MAX_iterations_H' parameter is $8$. If the condition ~(\ref{eq:Hmassvar}) is never verified, the new track will be the one with the minimum value of $\frac{\left|M_1-M_0\right|}{M_0}$.

\subsubsection{Stars in the He phase}
\label{subsubsec:hestars}
For this category we need to distinguish stars that have a Hydrogen envelope from WR stars.

i) To change track for WR stars, we use the look-up tables of bare Helium cores (see Sec. 2.1.2). To find a new track we use the same algorithm described for the stars in the H phase, with the following appropriate changes: $M_\mathrm{ZAMS}\rightarrow M_\mathrm{He-ZAMS}$, $\Theta_{\rm H}\rightarrow\Theta_{\rm He}\equiv\frac{t_{\mathrm{loc}}}{t_{\mathrm{COS}}}$, where $t_\mathrm{COS}$ is the time when the CO core starts to decouple from Helium.

ii) A new track for He stars with a Hydrogen envelope is successfully found if the following conditions are simultaneously verified 

\begin{equation}
\frac{\left|M_{\rm He,1}-M_{\rm He,0}\right|}{M_{\rm He,0}}<\epsilon_2,
\label{eq:WRmassvar}
\end{equation}

\begin{equation}
\frac{\left|M_1-M_0\right|}{M_0}<\epsilon_3,
\label{eq:HpHemassvar}
\end{equation}
where $M_{\rm He,1}$ is the mass of the He core of the new star, $M_{\rm He,0}$ is the mass of the He core of the old star, and $\epsilon_2$ and $\epsilon_3$ are two parameters with a typical value of $\sim 10^{-3}$ and $\sim 10^{-2}$, respectively. The algorithm starts to inspect the track with ZAMS mass given by equation ~(\ref{eq:firsttrack}) and the ZAMS is changed iteratively with a mass step $\Delta{}m=\pm{}0.2\,{}$. Here, we adopt the minus sign for a donor star (in this case we check for a new track with $M_{\mathrm{ZAMS}}<M_{\mathrm{ZAMS,}0}$) and the plus sign for an accretor (in this case we check for a new track with $M_{\mathrm{ZAMS}}>M_{\mathrm{ZAMS,}0}$).
The new star is searched in the time interval $\left[t_{\mathrm{0,He}},t_{\mathrm{f,He}}\right]$. If both the conditions ~(\ref{eq:WRmassvar}) and ~(\ref{eq:HpHemassvar}) are verified, a new track is found, otherwise the algorithm does not change the track.

\subsubsection{Stars in the CO phase}
In this case we use an algorithm analogous to that described for He stars (see section~\ref{subsubsec:hestars}). The difference is that for stars in the CO phase we search the new track in the time interval $\left[t_{\mathrm{0, CO}},t_{\mathrm{SN}}\right]$ where $t_{\mathrm{0, CO}}$ is the time when the CO core starts to decouple from Helium and $t_{\mathrm{SN}}$ is the time when the star transforms into a compact remnant. If the star is a WR, we search the new track in the time interval $\left[\mathrm{max}\left(t_{\mathrm{0, CO}},t_{\mathrm{He,max}}\right), t_{\mathrm{SN}}\right]$.

\subsection{The temporal evolution of a star}
To evolve the mass of a star from time $t_1$ to time $t_2=t_1+\Delta t$, we use the formula
\begin{equation}
\begin{split}
\label{eq:fakestar}
&M_2 = M_1 + V_m\,{}M_1,\\
&{\rm where}\,{}V_m=\frac{m_2-m_1}{m_1}. 
\end{split}
\end{equation}
In the above formula $M_2$ is the mass of the star at time $t_2$, $M_1$ is the mass of the star at time $t_1$, $m_1$ and $m_2$ are the masses of the star obtained from the interpolation tracks at time $t_1$ and $t_2$, respectively (see Equation \ref{eq:masscalc}), and $V_m$ is the relative variation of the mass of the star, calculated from the interpolation tracks. We use equation \ref{eq:fakestar} because, should the track-finding algorithm not converge (i.e. $\left|m_1-M_1\right|>\epsilon_1 M_1$, see equation \ref{eq:Hmassvar}), the temporal evolution of $M$ is still continuous.

In contrast, if the track-finding algorithm converges, we have $M_1\simeq m_1$, that is $M_2\simeq m_2$, which means that the evolution of the star is synchronous with the values in the look-up tables. We adopt the same technique for the temporal evolution of $M_\mathrm{He}$, and $M_\mathrm{CO}$, while we keep $R$, $L$ and time always synchronous with the values obtained with the interpolation tracks.

\section{Prescriptions for SNe} \label{app:sne}
Uncertainties on models of core-collapse SNe are still large (see \citealt{foglizzo2015} for a recent review). Overall, there is consensus that the properties of the progenitor star at the onset of core collapse determine the mass of the compact remnant, but the details differ significantly from one model to the other. For this reason, in {\sc SEVN} we decided to implement several different models, which can be activated with a different option in the parameter file. The models currently available in {\sc SEVN} are the following:
\begin{itemize}
	\item{} The delayed core-collapse model is described in \cite{fryer2012} and in \cite{spera2015}. It is based on the calculations by \cite{fryer2012} and on the idea that the shock is launched $>0.5$~s after the onset of core collapse. In this model, the final mass of the remnant depends just on the mass of the CO core and on the final mass of the star (i.e. the total mass before the onset of core collapse).
	
	\item{} The rapid core-collapse model is also described in \cite{fryer2012} and in \cite{spera2015}. The only difference between the rapid and the delayed SN model is the time when the shock is launched: $<250$ ms after the onset of core collapse in the case of the rapid SN model. Both the delayed and the rapid model depend only on the mass of the CO core and on the final mass of the progenitor star.

	\item{} The {\sc startrack} model is the same as adopted in the  {\sc startrack} code \citep{belczynski2010}. Also in this case, the mass of the final remnant depends only on the CO core mass and on the final mass of the star. The final remnant masses are similar to the ones obtained with the rapid model.
	
	\item{} The compactness model is based on the compactness of the stellar interior at the onset of core collapse, defined as \citep{oconnor2011}
	\begin{equation}
	\xi{}_{2.5}=\frac{2.5\,{}{\rm M}_\odot{}}{R(2.5\,{}{\rm M}_\odot{})/{\rm km}},
	\end{equation}
	which is the measure of a characteristic mass (in this case $2.5\,{}{\rm M}_\odot{}$) divided by the radius which encloses this mass at the onset of core collapse. Previous work \citep{oconnor2011,ugliano2012,horiuchi2014} shows that if $\xi{}_{2.5}\gtrsim{}0.2$ a star is expected to directly collapse to a BH. Unlike the former three models, the model based on the compactness requires that we know the internal properties of a star at the onset of collapse, when a Fe core is already formed. Thus, this model cannot be used self-consistently in combination with stellar-evolution models that do not describe nuclear burning up to the formation of a Fe core.
	
	\item{} The two-parameter model is based on the mass for which the dimensionless entropy per nucleon is $s=4$ ($M_4$) and on the mass gradient at the same location ($\mu{}_4=dM/dr|_{s=4}$). \cite{ertl2016} have proposed this model, based on the fact that the complex physics of core-collapse SNe cannot be described entirely by a single parameter like the compactness. The underlying idea is that $\mu_4$ scales with the ram-pressure on the infalling material from the outer layers of the collapsing star, while $M_4\,{}\mu_4$ scales with the neutrino luminosity. As for the compactness, also the two-parameter model requires that we know the internal properties of a star at the onset of core collapse.
	
\end{itemize}
The first three models we described (delayed, rapid and {\sc startrack}) depend only on the CO core mass and on the final mass of a star, while the latter two models depend on the internal structure of a star at the onset of core collapse. The latter models are more accurate but require modelling the interior structure of a star at the onset of core collapse. Recently, \cite{limongi2017} and \cite{limongi2018} have shown that there is a strong correlation between the CO core mass and the compactness at the onset of core collapse, suggesting that even the more approximated models capture the main features of core collapse.

Unlike core-collapse SNe, the physical mechanisms powering PISNe have been understood and satisfactorily described \citep{ober1983, bond1984, heger2003, woosley2007}. In very massive, metal-poor stars the central temperature can rise above $\sim{}7\times{}10^8$~K, leading to an effective production of positron and electron pairs. This removes radiation pressure, causing the core to contract. The result is an increase of the central temperature, leading to an early and simultaneous switching on of Oxygen and Silicon burning. If $64\lesssim{}M_{\rm He}/{\rm M}_\odot\lesssim{}135$ (where $M_{\rm He}$ is the Helium core mass) the star is completely destroyed by the explosive burning of Oxygen and Silicon \citep{woosley2017}, leaving no compact remnant. This mechanism is known as PISN. If $M_{\rm He}>135$ M$_\odot$, the early contraction of the core cannot be stopped and the star collapses to a BH directly. If $32\lesssim{}M_{\rm He}/{\rm M}_\odot\lesssim{}64$, the stellar core can undergo one or more oscillations, during which mass loss is significantly enhanced. This mechanism is known as pulsational PISN (PPISN, \citealt{woosley2017}). At the end of the oscillations, the star finds a new equilibrium and dies with a core-collapse SN. As described in \cite{spera2017}, we have included both PISNe and PPISNe in {\sc SEVN}, following the models of \cite{woosley2017}. In particular, the mass of the remnant is described as
\begin{equation}
m_{\rm rem, PISN}=f(M_{\rm He},\,{}m_{\rm fin})\,{}m_{\rm rem, no PISN},
\end{equation}

where $m_{\rm rem, PISN}$ ($m_{\rm rem, no PISN}$) is the final mass of the compact remnant when we account (we do not account) for PISNe and PPISNe, while $f(M_{\rm He},\,{}m_{\rm fin})$ is a function of the Helium core mass and of the final mass of the star, described in Appendix~B of \cite{spera2017}.

Another issue related to SNe is the natal kick of the compact remnant. There are no direct measurements of the natal kick of BHs, but only indirect studies based on the proper motion of few X-ray binaries \citep{gualandris2005,fragos2009,repetto2012,repetto2017}. As for neutron stars (NSs), \cite{hobbs2005} have derived the proper motions of 233 isolated pulsars in the Milky Way, showing that their distribution can be fit with a Maxwellian distribution with one-dimensional root-mean square $\sigma{}_{\rm kick}=265$ km s$^{-1}$. This result is still debated (e.g. \citealt{fauchergiguere2006,verbunt2017}), especially for binary NSs \citep{beniamini2016,beniamini2016b,giacobbo2018c}, but is still the most used distribution for natal kicks of NSs.

In \sevn{}, we adopt the \cite{hobbs2005} kick distribution for both NSs and BHs but
we scale it by the amount of fallback \citep{fryer2012}:
\begin{equation}
V_{\rm kick} = (1 - f_{\rm fb})\,{}W_{\rm kick},
\end{equation}
where $f_{\rm fb}$ is the fallback factor (the explicit expression can be found in \citealt{giacobbo2018}), and $W_{\rm kick}$ is randomly drawn from the Maxwellian distribution derived by   \cite{hobbs2005}. According to this formalism, if a BH forms by prompt collapse of the parent star $V_{\rm kick}=0$.

If the SN occurs when the BH or NS progenitor is member of a binary, the SN kick can unbind the system. The survival of the binary system depends on the orbital elements at the moment of the explosion and on the SN kick. If the binary remains bound, its post-SN semi-major axis and eccentricity are calculated as described in appendix A1 of \cite{hurley2002}.

\section{Mass transfer}\label{app:masstransfer}
\subsection{Wind mass transfer}
The mean accretion rate by stellar winds is calculated
as (\citealt{bondi1944}, see also eq.~6 of \citealt{hurley2002})
\begin{equation}\label{eq:windacc}
\langle{}\dot{M}_{2}\rangle{}=\frac{1}{\sqrt{1-e^2}}\,{}\left(\frac{G\,{}M_{2}}{v^2_{\rm W}}\right)^2\,{}\frac{\alpha{}_{\rm W}}{2\,{}a^2}\,{}\frac{1}{(1+v^2)^{3/2}}\,{}\dot{M}_{1},
\end{equation}    
where $\dot{M}_1$ is the mass lost by the donor by stellar winds ($\dot{M}_1>0$), $M_2$ is the accretor mass, $e$ is the orbital eccentricity, $G$ is the gravitational constant, $a$ is the semi-major axis, and $\alpha{}_{\rm W}=1.5$ \citep{hurley2002}, while $v$ and $v_{\rm W}$ are defined as follows: 
\begin{eqnarray}
v^2=\frac{G\,{}(M_1+M_2)}{a\,{}v^2_{\rm W}},\\
v^2_{\rm W}=2\,{}\beta{}_{\rm W}\left(\frac{G\,{}M_1}{R_1}\right),
\end{eqnarray}
where $M_1$ is the mass of the donor, $R_1$ is the radius of the donor and the dimensionless parameter $\beta_{\rm W}\sim{}0.1-7$ depends on the spectral type \citep{hurley2002}.  $\langle{}\dot{M}_2\rangle{}$ in equation~\ref{eq:windacc} is averaged over an orbital period and is strictly valid only if $v_{\rm W}>>G\,{}(M_1+M_2)/a$. We impose that $\dot{M}_2\leq{}0.8\,{}|\dot{M}_1|$ to avoid that more mass is accreted by the secondary than is lost by the primary, under some special circumstances.

Non-conservative mass transfer also induces a change in the angular momentum of the system. Following \cite{hurley2002}, we describe the orbit-averaged change of angular momentum due to wind mass transfer as
\begin{equation}\label{eq:angmom}
\dot{J}_{\rm orb}=\left(\dot{M}_1\,{}M_2 - M_1\,{}\dot{M}_2\right)\,{}M_2\,{}(1-e^2)^{1/2}\,{}\left[\frac{G\,{}a}{(M_1+M_2)^3}\right]^{1/2}
\end{equation}
In  equation~\ref{eq:angmom}, we assume that only the primary loses mass and only the secondary accretes mass, which is not true in the general case, because both stars lose mass by stellar winds. If we assume that both binary members donate and accrete mass at the same time, equation \ref{eq:angmom} is generalized as
\begin{eqnarray}\label{eq:angmom2}
\dot{J}_{\rm orb}=\left[\left(\dot{M}_{\rm 1L}\,{}M_2- M_1\,{}\dot{M}_{\rm 2A}\right)\,{}M_2+\left(\dot{M}_{\rm 2L}\,{}M_1- M_2\,{}\dot{M}_{\rm 1A}\right)\,{}M_1\right]\nonumber{}\\\,{}(1-e^2)^{1/2}\,{}\left[\frac{G\,{}a}{(M_1+M_2)^3}\right]^{1/2},
\end{eqnarray}
where $\dot{M}_{\rm 1L}$ and $\dot{M}_{\rm 1A}$ ($\dot{M}_{\rm 2L}$ and $\dot{M}_{\rm 2A}$) are the mass loss rate and the mass accretion rate of the primary (secondary), respectively\footnote{The sign of equations~\ref{eq:angmom} and \ref{eq:angmom2} is different from the one reported by \cite{hurley2002} only because in our formalism the mass loss rate is positive $\dot{M}_{\rm 1L}>0$, $\dot{M}_{\rm 2L}>0$.}.

Non-conservative wind mass transfer also affects the spins of the stars. The change of the spin angular momentum of the primary due to stellar winds is described as  \citep{hurley2002}
\begin{equation}
\dot{J}_{\rm spin,1}=-\frac{2}{3}\dot{M}_{\rm 1L}\,{}R_1^2\,{}J_{\rm spin,1}\,{}I_1^{-1}+\frac{2}{3}\dot{M}_{\rm 1A}\,{}R_2^2\,{}J_{\rm spin,2}\,{}I_2^{-1},
\end{equation}
where $\dot{M}_{\rm 1L}$ is the mass loss rate by stellar winds of the primary, $\dot{M}_{\rm 1A}$ is the mass accretion rate by wind accretion of the primary, $J_{\rm spin,1}$ ($J_{\rm spin,2}$) is the spin angular momentum of the primary (secondary) and $I_1$ ($I_2$) is the inertia of the primary (secondary). The change of the spin angular momentum of the secondary is described in the same way, by changing the subscripts accordingly. 


Following \cite{hurley2002}, the orbit-averaged change of eccentricity is 
\begin{equation}
\frac{\dot{e}}{e}=-\dot{M}_2\,{}\left(\frac{1}{M_1+M_2}+\frac{1}{2\,{}M_2}\right),
\end{equation}
where $\dot{M}_2$ is the mass accretion rate averaged over a time-step.

\subsection{Roche-lobe overflow}
At every time-step we evaluate whether one of the two members of the binary fills its Roche lobe by using equation 6 \citep{eggleton1983}.

If the Roche-lobe filling donor is a neutron star (NS, $k=13$) or a BH ($k=14$), the accretor must be another NS or BH. In this case, the two objects are always merged. In all the other cases, to decide the amount of mass transferred from the primary $\Delta{}m_1$, we first evaluate the stability of mass transfer using the radius-mass exponents $\zeta{}$ defined by \cite{webbink1985}. In particular, $\zeta{}_{\rm ad}\equiv{}\frac{d\ln{R_1}}{d\ln{m_1}}|_{\rm ad}$ is the change of radius of the donor needed to reach a new hydrostatic equilibrium as a consequence of mass loss, $\zeta{}_{\rm th}\equiv{}\frac{d\ln{R_1}}{d\ln{M_1}}|_{\rm th}$ is the change of radius of the donor needed to reach a new thermal equilibrium as a consequence of mass loss, and $\zeta{}_{\rm L}\equiv{}\frac{d\ln{R_{L,1}}}{dM_1}$ is the change of the Roche lobe induced by mass loss. $\zeta{}_{\rm ad}$, $\zeta{}_{\rm th}$ and $\zeta{}_{\rm L}$ are calculated as described in \cite{hurley2002}.

Following \cite{hurley2002} we do not estimate $\zeta_{\rm ad}$ directly, but we adopt a simplified criterion: we use the critical mass ratio $q_{\rm c}$ defined as the mass ratio for which $\zeta{}_{\rm ad}=\zeta_{\rm L}$ \citep{soberman1997}. Following {\sc BSE},
\begin{equation}
q_{\rm c}=\begin{cases}0.695 & \textrm{if } k = 0 \cr
3 & \textrm{if } k = 1, 4 \cr
4 & \textrm{if } k = 2 \cr
0.362 + 1.0\,{}\left[3.0\,{}(1.0 - M_{\rm c,1}/M_1)\right]^{-1} & \textrm{if } k = 3, 5, 6 \cr
0.784 & \textrm{if } k = 8, 9 \cr
0.628 & \textrm{if } k = 10, 11, 12 \cr
\end{cases}
\end{equation}
where $M_{\rm c,1}$ is the mass of the core of the donor\footnote{{ Up-to-date values for $\zeta_{\rm ad}$ and $q_{\rm c}$ can be found in \citet{claeys2014} and in \citet{ge2015}, respectively. We will include the new values in the next version of the \sevn{} code.}}. 

If the Roche-lobe filling donor is a  first giant branch star (type $k=3$) or a core Helium burning (cHeB) star ($k=4$) or an asymptotic giant branch (AGB) star ($k=5,6$) or an Hertzsprung gap (HG) Naked Helium star ($k=8$) or a Giant Branch Naked Helium star ($k=9$), we start a CE phase if $q_1>q_C$.

If a deeply convective MS star (mass $<0.7$ M$_\odot$, $k=0$) fills its Roche lobe and $q_1>q_c$, the two stars are merged. The mass accreted by the merger product is decided as in \cite{hurley2002}.

If both the donor and the accretor are  MS  stars with mass $>0.7$ M$_\odot$ ($k=1$) or HG stars ($k=2$) and $q_1>q_c$, we always merge them (note that \cite{hurley2002} allow HG stars to enter CE rather than being merged if $q_1>q_c$).

If the donor is a WD ($k=10,11,12$) and $q_1>q_c$, the two stars are merged. The treatment of the merger product is the same as described in \cite{hurley2002}.  The formalism based on $q_{\rm c}$ contains several simplifications. In the future updates of {\sc SEVN} we will include a more accurate formalism.

If $\zeta_{L}>\zeta_{\rm ad}$, mass transfer is unstable over a dynamical timescale (i.e. the radius of the primary increases faster than the Roche lobe on conservative mass transfer). If this condition is satisfied when the donor is a MS or a HG star, the binary is merged. If this condition is satisfied by any other non-degenerate donor,  the binary enters a CE phase.

If $\zeta{}_{\rm L}<(\zeta_{\rm ad},\zeta_{\rm th})$ mass transfer is stable, until nuclear evolution changes the radius of the star. In this case, the mass loss rate of the primary is described by equation 7.

If $\zeta_{\rm th}<\zeta_{\rm L}<\zeta_{\rm ad}$, mass transfer is unstable on a thermal timescale. Equation 7 can be considered an upper limit to mass loss in this case, because the thermal timescale is small compared to the nuclear timescale. We thus calculate the mass loss as the minimum between the values given in equation 7 and 8.

The accreted mass $\Delta{}m_2$  in the case of a stable mass transfer or of a thermally unstable mass transfer is described by equation 9 if the accretor is not a compact object, and by equation 10 if the accretor is a compact object.

If the accretor is a WD we also consider the possibility of a nova eruption. In particular, if the donor is Hydrogen rich ($k\leq{}6$) and $\dot{M}_1<1.03\times{}10^{-7}$ M$_\odot$ yr$^{-1}$, we assume that a nova occurs and the accreted matter is only $\Delta{}m_{\rm 2,nova}=f_{\rm nova}\,{}\Delta{}m_2$, where $f_{\rm nova}=0.001$ \citep{hurley2002}.



Non-conservative mass transfer also affects the orbital angular momentum of the system and the spins of the star. The variation of orbital angular momentum is described as
\begin{equation}
\dot{J}_{\rm orb}=(\dot{M}_1-\dot{M}_2)\,M_2^2\,{}(1-e^2)^{1/2}\,{}\left[\frac{G\,{}a}{(M_1+M_2)^3}\right]^{1/2}
\end{equation}
In this equation, we assume that the material lost from the system carries with it the specific angular momentum of the primary.

If the accretion onto a compact object is super-Eddington or if there is a nova eruption, we use a different prescription for the variation of the orbital angular momentum:
\begin{equation}
\dot{J}_{\rm orb} = (\dot{M}_1 - \dot{M}_2)\,{}M_1^2\,{}(1-e^2)^{1/2}\,{}\left[\frac{G\,{}a}{(M_1+M_2)^3}\right]^{1/2},
\end{equation}
which means that we assume that this mass is lost by the system as a wind  from the secondary.


The loss of spin angular momentum of the primary by Roche lobe overflow is described as
\begin{equation}
\dot{J}_{\rm spin,1}= \dot{M}_1\,{}R_{\rm L,1}^2\,{}J_{\rm spin,1}\,{}I_1^{-1},
\end{equation}
where $R_{\rm L,1}$ is the Roche lobe of the primary, while $J_{\rm spin,1}$ and $I_1$ are the spin angular momentum and the inertia of the primary (i.e. the Roche lobe filling star).

The spin up of the secondary (i.e. the accretor) depends on whether an accretion disc forms around it. According to {\sc BSE}, the accretion disc radius is estimated as \citep{ulrich1976}
\begin{equation}
R_{\rm D}= 0.0425\,{}{\rm R}_\odot\,{}a\,{}[q_2\,{}(1+q_2)]^{1/4}
\end{equation}
If $R_{\rm D}>R_2$, then a disc forms and the change of spin angular momentum of the secondary is
\begin{equation}
\dot{J}_{\rm spin,2}= \dot{M}_2\,{}(G\,{}M_2\,{}R_{\rm L,2})^{1/2},
\end{equation}
where $R_{\rm L,2}$ is the Roche lobe of the accretor. Here we assume that material falls onto the star from the inner edge of a Keplerian accretion disc and that the system is in a steady state. 

If $R_{\rm D}\leq{}R_2$, then we calculate the change of spin as
\begin{equation}
\dot{J}_{\rm spin,2}= \dot{M}_2\,{}(G\,{}M_2\,{}1.7\,{}R_{\rm D})^{1/2}.
\end{equation}

This results in a spin up of the accretor. We then check if the final spin is larger than the critical spin (above which the star is expected to break up)
\begin{equation}
J_{\rm crit}= I_2\,{}\left(\frac{2}{3}\right)^{3/2}\,{}\left[\frac{G\,{}(M_2+\Delta{}m_2)}{R_2^3}\right]^{1/2},
\end{equation}
where $I_2$ is the inertia of the accretor. 

If $(J_{\rm spin,2}+\dot{J}_{\rm spin,2}\,{}dt)>J_{\rm crit}$, we force the final spin of the accretor to be the same as $J_{\rm crit}$. 
It is not clear whether stars can keep accreting once they reach the break-up spin (see \citealt{packet1981}, \citealt{popham1991}, \citealt{petrovic2005}, \citealt{demink2013}). Here we assume that viscous coupling with the circumstellar disk can efficiently remove angular momentum from the star without halting the accretion flow.

\section{Supplementary figures}
\begin{figure*}
	\includegraphics[width=0.9\hsize]{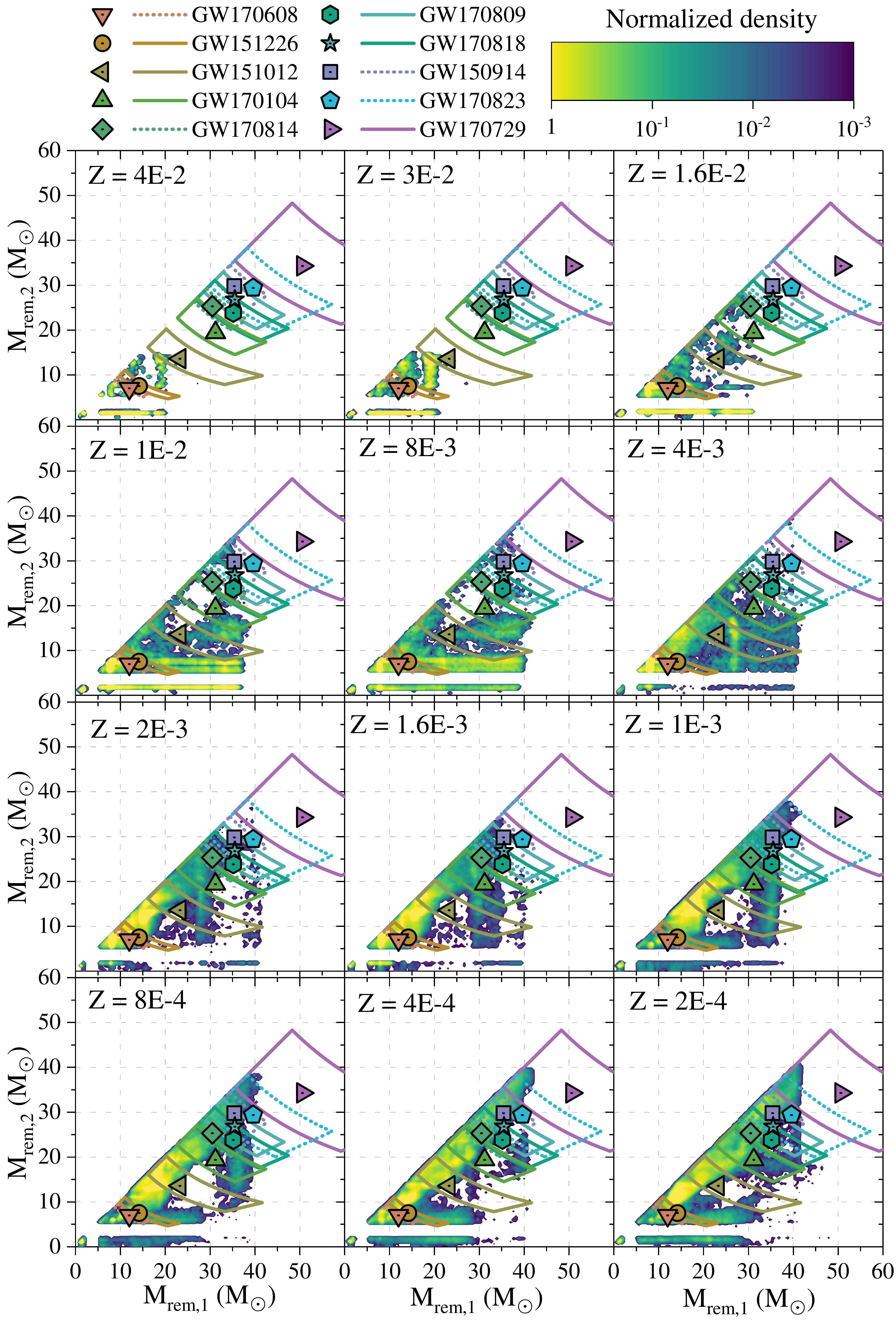}
	\caption{Same as the panels in the bottom row of Fig. 6, but for all the other metallicities considered in our simulations.}
	\label{fig:allz_merging}
\end{figure*}

\begin{figure}
	\includegraphics[width=\hsize]{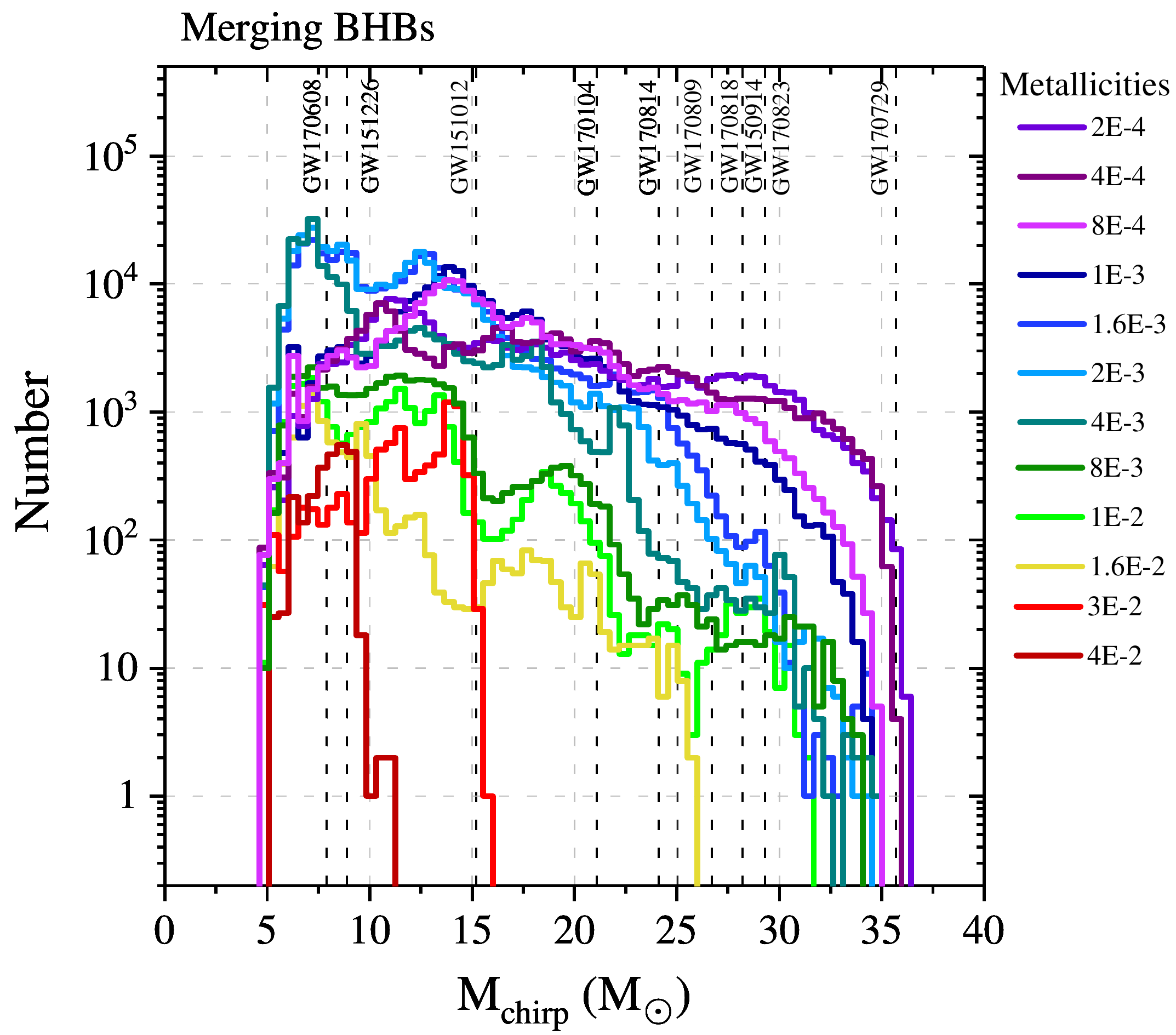}
	\caption{Same as Fig. 7, but for all the considered metallicities.}
	\label{fig:allz_mchirp}
\end{figure}

\begin{figure}
	\includegraphics[width=\hsize]{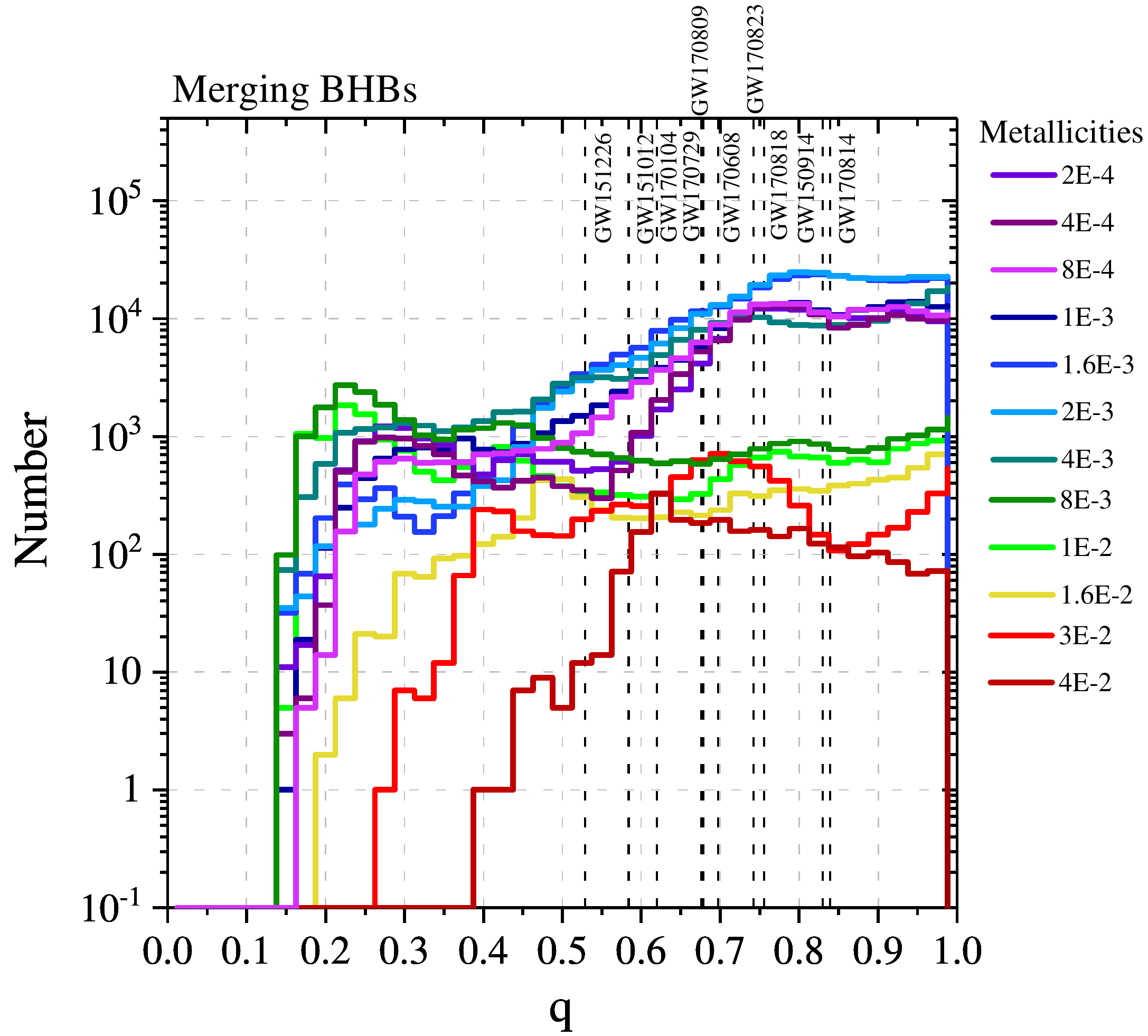}
	\caption{Same as Fig. 8, but for all the considered metallicities.}
	\label{fig:allz_q}
\end{figure}

In this appendix we show the masses of merging compact-object binaries (Fig. \ref{fig:allz_merging}), the distribution of chirp masses of merging BHBs (Fig. \ref{fig:allz_mchirp}) and the distribution of the mass ratio of merging BHBs (Fig. \ref{fig:allz_q}) for all the considered metallicities ($Z\in \left[10^{-4};4\times 10^{-2}\right]$).

\bibliography{./bibliography}

\end{document}